\documentclass[reprint,twocolumn,aps,prb,superscriptaddress]{revtex4-1}
\usepackage[T1]{fontenc}
\usepackage[utf8]{inputenc}
\usepackage{lmodern}
\usepackage{graphicx}
\graphicspath{{figures/}{.}}
\usepackage{amsmath}
\usepackage{xcolor}\usepackage{xcolor}
\usepackage[normalem]{ulem}

\usepackage{float}
\usepackage{hyperref}
\usepackage{braket}
\usepackage{amssymb,amsmath,amsthm, array}
\usepackage{mdframed}
\usepackage{systeme,mathtools}

\DeclareUnicodeCharacter{0301}{-}

\begin{document}
\renewcommand{\vec}[1]{\mathbf{#1}}
\newcommand{\ii}{\mathrm{i}}
\def\ya#1{{\color{orange}{#1}}}

\title{Quantum discord and entropic measures of two relativistic fermions}

\author{Podist Kurashvili}
\affiliation{National Centre for Nuclear Research, Warsaw 00-681, Poland}

%

\author{Levan Chotorlishvili}
\affiliation{Department of Physics and Medical Engineering, Rzesz\'ow University of Technology, 35-959 Rzesz\'ow, Poland}

\date{\today}
\begin{abstract}
In the present work, we study the interplay between 
relativistic effects and quantumness in the system of two relativistic fermions. 
In particular, we explore entropic measures of quantum 
correlations and quantum discord before and after
application of a boost and subsequent Wigner rotation.
We also study the positive operator-valued measurements (POVM) 
invasiveness before and after the boosts. 
While the relativistic principle is universal and requires Lorentz 
invariance of quantum correlations in the entire system, we have 
found specific partitions where quantum correlations stored in particular 
subsystems are not invariant. 
We calculate quantum discords corresponding of the 
states before and after applying a boost, and observe that 
the state gains extra discord after the boost.
When analyzing the invasiveness of the POVMs, we have found that the POVM 
applied to the initial entangled state reduces the discord to zero. 
However, discord of the boosted state survives after the same POVM.
Thus we conclude that the quantum discord generated by Lorentz boost is 
robust concerning the protective POVM, while the measurement exerts an 
invasive effect on the discord of the initial state. 
Finally, we discuss potential implementation of the ideas of this work 
using top quarks as a benchmark scenario.
\end{abstract}


\maketitle

\tableofcontents

\section{Introduction}
\label{sec:General}
Entanglement is a central concept of quantum information theory. Interest in quantum entanglement has fundamental theoretical, academic, and practical aspects. Entanglement can be utilized in quantum communication and quantum computations. The quantum information literature is dominantly focused on condensed matter physics, mainly non-relativistic systems
\cite{PhysRevLett.80.2245,PhysRevLett.86.3658,wilde2013quantum,PhysRevB.100.174413,chotorlishvili2011thermal,PhysRevB.96.054440}. Nevertheless, last period attention was paid to the entanglement in a relativistic realm 
\cite{casini2009entanglement,nishioka2009holographic,PhysRevA.80.044302,PhysRevA.81.062320,PhysRevA.84.012334,PhysRevD.97.016011,PhysRevLett.111.090504,kurashvili2021coherence,PhysRevD.103.036011,Afik:2020onf,Afik:2022kwm,Severi:2021cnj,kurashvili2022quantum}. Among different signatures of quantum entanglement, a central role belongs to entropic measures. In opposite to classical systems, quantum entropies are not additive quantities. For instance, while the entropy of the entire bipartite pure state $\hat \varrho_{AB}$ is zero $S\left(\hat \varrho_{AB}\right)=-Tr\left(\hat \varrho_{AB}\ln\hat \varrho_{AB}\right)=0$, typically the entropy of each subsystem $\hat \varrho_{A}=Tr_B\left(\hat \varrho_{AB}\right),~\hat \varrho_{B}=Tr_A\left(\hat \varrho_{AB}\right)$ is nonzero $S\left(\hat \varrho_{A}\right)=S\left(\hat \varrho_{B}\right)\neq 0$. In essence, nonzero entropy in the subsystems is a signature of quantum entanglement and underlines the difference between the quantum and the classical world. 

The relativistic character of a system may play an intriguing role in the partitioning of the Hilbert space when calculating the entropies of subsystems. Invariance of the entire system or its subsystem under the Lorentz boosts is required according to the fundamental physical principle and holds rigorously. Entropies of subsystems are invariant, $S\left(\hat \varrho_{A}\right)=S\left(\hat \varrho_{A}^\Lambda\right)=S\left(\hat \varrho_{B}\right)=S\left(\hat \varrho_{B}^\Lambda\right)=\mathrm{const}$, where $\hat \varrho_{A}^\Lambda,\,\hat \varrho_{B}^\Lambda$ and $\hat \varrho_{A},\,\hat \varrho_{B}$ are reduced density matrices of subsystems $A$ and $B$ in the boosted and rest frames, respectively. However, invariance of the entropy under Lorentz transformations can be violated for another type of partitioning \cite{PhysRevA.81.042114}. Suppose that both subsystems $A$ and $B$ of the bipartite state $\hat \varrho_{AB}$ are in turn composed by two subsystems, e.g., spin and momentum sectors $\hat \varrho_{AB}=\hat \varrho_{\sigma_Ap_A\sigma_Bp_B}$.
In the c.m. frame, we can bipartite the entire system as $\hat \varrho_{AB}=\hat \varrho_{\sigma_A p_A}\otimes\hat \varrho_{\sigma_B p_B}$,    $\sigma=(\hat\sigma_A,\,\hat\sigma_B)$, $P=(p_A,\,p_B)$, and explore entanglement between spin and momentum sectors. There is no principle bound on the conservation of subsectors entanglement. Therefore, through Lorenz boosts, we can reshuffle entanglements of different sectors. We analyse this problem for a relativistic quantum system of two fermions. The results discussed here are quite general and concern different fields such as quantum chromodynamics (QCD), electroweak theory, and high-energy physics in general. 

We consider the entanglement between the spins and momenta of two relativistic massive spin-$1/2$ particles. The mathematical formalism and derivations are simpler in the c.m. frame, where the particles have equal and antiparallel momenta. The spin state can be ether singlet or triplet. Without loss of generality, we adopt the singlet state for illustrative purposes. In the c.m. frame of the particles, we consider the total quantum state of the system as a trivial product of the momentum and spin wave functions. The separability breaks after we change the reference frame by a boost perpendicular to the direction of the momenta. After the boost, the spin and momentum degrees of freedom get entangled, and extra boost-dependent entanglement is added to the total entanglement of the system. 

The general formalism presented here is of particular relevance for the study of particle-antiparticle production at high-energy colliders.
In this environment, massive fermions are created typically in pairs of particle and antiparticle with relativistic momentum.
The momentum of those fermions is measured, and spin related properties, such as entanglement, can be deduced by angular distributions of the measured objects. 
We discuss the special case of a pair of top and antitop quarks generated at a hadron collider, as it was recently shown that entanglement in this specific scenario can be measured~\cite{Afik:2020onf,Afik:2022kwm}.

The work is structured as follows: In Section \textbf{\ref{subsec:Entropic measures}} we specify the entropic measures used afterwards. In Section \textbf{\ref{sec:particleantiparticle}} we describe the quantum formalism of particle-antiparticle production. In Section \textbf{\ref{sec:WaveFunction}} we
calculate the total entanglement of the particle-antiparticle system, exploiting the c.m. frame and different
partitions of Hilbert space. The separable Hilbert space comprises the states of two spins and two momentum variables. 
We calculate the quantum discord and entropic measures of the post-measurement density matrix after measuring the spin of one of the particles.
We use the results of calculations to find the relative information between the spins and momenta.
In Section \textbf{\ref{sec:Relativistic}} we perform the same calculations for the
boosted state. The non-trivial effects of the boosts depend on the angular parameter characterizing the boost (to be defined later). In Section \textbf{\ref{sec:Discord}} we calculate the quantum discord between the two spins in the rest frame and after the boost. Finally, in Sec. \textbf{\ref{sec:Experiment}} we discuss potential experimental implementations of the results of this work using top quarks. Technical details and derivations we present in the Appendix.

\section{Entropic measures}
\label{subsec:Entropic measures}

Below we define the mathematical formalism and quantities used hereafter. We first review the basics of quantum measurements. Consider the quantum state of a certain system that is described by a pure state
\begin{equation}
    \ket{\Psi}=\sum_n c_n\ket{\phi_n}
\end{equation}
where the states $\ket{\phi_n}$ characterize the eigenstates of some observable $O$ with eigenvalues $O_n$.
We consider a simple model of pre-measurement quantum state, in which the system couples to the apparatus measuring $O$, described by states $\ket{\chi_n}$. This gives a total quantum state System+Apparatus of the form
\begin{equation}
    \ket{\Psi_\textrm{T}}=\sum_n c_n\ket{\phi_n}\otimes\ket{\chi_n}
\end{equation}
This total quantum state is described by the density matrix
\begin{equation}
    \hat\varrho_{\textrm{T}}=\ket{\Psi_\textrm{T}}\bra{\Psi_\textrm{T}}=\sum_{n,m} c_nc^*_m\ket{\phi_n}\otimes\ket{\chi_n}\bra{\phi_m}\otimes\bra{\chi_m}
\end{equation}
After tracing over the external degrees of freedom of the detector, we get the reduced density matrix describing only the quantum state of the system
\begin{equation}
    \hat\varrho_O =\textrm{Tr}_A\hat\varrho_\textrm{T}=\sum_n|c_n|^2\ket{\phi_n}\bra{\phi_n}
\end{equation}
The mixed state $\hat\varrho_O$ contains all the information about $\ket{\Psi}$ accessible in an experiment in which we measure $O$.

The same result could have been derived by using the formalism of positive operator-valued measurements (POVM), where the post-measurement density matrix $\hat\varrho_O$ is directly obtained after projecting onto the relevant states
\begin{equation}
    \hat\varrho_O=\sum_n \Pi_n\ket{\Psi}\bra{\Psi}\Pi_n=\sum_n|c_n|^2\ket{\phi_n}\bra{\phi_n}
\end{equation}
with $\Pi_n$ is the rank-1 POVM projection operator onto the eigenstate $\ket{\phi_n}$, $\Pi_n=\ket{\phi_n}\bra{\phi_n}$. If the measured states $\ket{\phi_n}$ do not span over the Hilbert space, the previous expression should have been properly normalized in order to ensure $\mathrm{Tr}\hat\varrho_O=1$,
\begin{equation}
    \hat\varrho_O=\frac{\sum_n \Pi_n\ket{\Psi}\bra{\Psi}\Pi_n}{\sum_n p_n},~p_n=\bra{\Psi}\Pi_n\ket{\Psi}
\end{equation}
with $p_n$ the probability of measuring $O_n$.

In the present work, we typically consider quantum states in a multipartite Hilbert space, composed of several subsystems. The linear entropy of the density matrix $\hat\varrho$
of the composite system corresponding
to a particular partition $P$ into several parts
is defined as the sum of entropies of different parts of the system:
\begin{equation}
E_P(\hat\varrho) = \sum_i \left(1-\mathrm{Tr}\hat\varrho_i^2\right),
\label{eq:Entanglement_definition}
\end{equation}
where $\hat\varrho_i$ denotes a reduced density matrix obtained by tracing over all
subsystems except of $i$-th subsystem, and the sum runs over all elements of the
partition. 
As it is shown in Appendix \ref{sec:CalculationOfEntropy}, given
a set of reduced density matrices $\hat\varrho_i$ with elements $\{\rho^i_{mn}\}$, Eq. (\ref{eq:Entanglement_definition}) reduces to the expression
\begin{equation}
E_P(\hat\varrho) = \sum_i (1 - \sum_{mn}\vert \rho^i_{mn}\vert^2).
\label{eq:Entanglement_definition_expanded}
\end{equation}
Linear entropy provides a quantitative measurement of the degree of entanglement while allowing the derivation of simple analytical formulae, in contrast the logarithmic Von Neumann entropy.

We consider POVMs done on both momentum and spin degrees of freedom throughout the work. After measuring the variable of the $k$-th subsystem, the state of the system is described by the post-measurement density matrix 
\begin{equation}
\hat \varrho_k = 
\sum_{i_k} \left(\ket{\varphi^k_{i_k}}\bra{\varphi^k_{i_k}} \otimes I_k\right)
\varrho \left(\ket{\varphi^k_{i_k}}\bra{\varphi^k_{i_k}} \otimes I_k \right),
\label{eq:DensityMatrix_AfterMeasurement}
\end{equation}
where $I_k$ denotes a unit matrix in the complementary space to the $k$-th subsystem,
and $\ket{\varphi^k_{i_k}}$ are the eigenstates of the measured variable corresponding to a value with index $i_k$.

The conditional entropy between two subsystems $X$ and $Y$ has the form:
\begin{equation}
E(X|Y) = E(X,Y)-E(Y)=E(\hat\varrho_{XY}) - E(\hat\varrho_Y),
\label{eq:Conditional_entropy_definition}
\end{equation}
and the mutual information between $X$ and $Y$ is defined as follows:
\begin{equation}
I(X, Y) = E(X) + E(Y) - E(X,Y).
\label{eq:Mutual_information_definition}
\end{equation}

Classically, an equivalent expression for the mutual information is
\begin{equation}
J(X,Y) = E(X)-E(X\vert Y).
\label{eq:Classical_correlation}
\end{equation}
However, for a quantum system, a quantum version of Eq. (\ref{eq:Classical_correlation_expand}) arises when considering a set of POVMs $\{\hat \Pi_i\}$ performed at the $Y$ subsystem, where $\hat \Pi_i=\vert i\rangle\langle i\vert$ is the POVM projector applied to $Y$, with $\ket{i}$ the corresponding eigenstate associated to the measurement. The resulting quantum expression for $J(X,Y)$ is calculated through the formula:
\begin{equation}
J(X, Y)_{\{\hat \Pi_i\}} = E(X)-E(X|\{\hat \Pi_i\}Y),
\label{eq:Classical_correlation_expand}
\end{equation}
where
\begin{equation}
    E(X|\{\hat \Pi_i\}Y) = \sum_i p_i E(\hat \varrho_{X\vert \{\hat \Pi_i\}Y})
\end{equation}
is the conditional entropy of the post-measurement state
\begin{equation}
\label{eq:POVMState}
\hat \varrho_{X\vert \{\hat \Pi_i\}Y} = \frac{\Pi_i \varrho_{XY} \Pi_i} {p_i},~p_i = \mathrm{Tr}\left(\Pi_i \hat \varrho_{XY}\right).
\end{equation}

Quantum discord is then defined through the minimum of the differences between the classical and quantum expressions for the mutual information:
\begin{eqnarray}
D(X,Y) &=& \min_{\{\hat\Pi_i\}} \left[I(X,Y) - J(X, Y)_{\{\hat\Pi_i\}}\right]\label{eq:Quantum_discord}\\
&=&\min_{\{\hat\Pi_i\}} \left[E(Y) -E(X,Y)+ E(X|\{\hat \Pi_i\}Y)\right].
\nonumber
\end{eqnarray}
In the following, we use entropic measures of quantum correlations and apply the above mathematical formalism to relativistic particle-antiparticle production.

\section{Quantum formalism in particle-antiparticle production}
\label{sec:particleantiparticle}




We describe here the formalism used to study the relativistic production of a particle-antiparticle pair \cite{Afik:2022kwm}. In what follows we adopt standard notations of quantum field theory. Greek indices $\mu,\,\nu$, run over the values $(0,1,2,3)$, and Einstein summation convention is oftenly understood unless otherwise stated. The metric is given through $g_{\mu\nu}=\textrm{diag}[1,-1,-1,-1]$. The four-momentum takes the form $p=(p^0,\mathbf{p})$, satisfying the usual Lorentz-invariant dispersion relation $p^\mu p_\mu=(p^0)^2-\mathbf{p}^2=m^2$, with $m$ the mass of the particle, and we adopt dimensionless units $\hbar=1$, $c=1$.

A particle-antiparticle pair typically arises in a relativistic process as the annihilation of some initial state $I$. Due to conservation of energy and momentum, the initial state $I$ has the same energy and total momentum that the produced particle-antiparticle pair. For the kinematical description of the production process, we switch to the c.m. frame, where the pair is described by its invariant mass $M$ and the direction of the particle $\hat{p}$. In this frame, the particle/antiparticle four-momenta are, respectively, $p_{\pm}^{\mu}=(p^0,\pm \mathbf{p})$, with $\hat{p}=\mathbf{p}/|\mathbf{p}|$. The invariant mass $M$ is the c.m. energy of the pair, defined from the usual invariant Mandelstam variable
\begin{equation}
    M^2\equiv s=(p_{+}+p_{-})^2,
\end{equation}
Regarding the quantum state of the particle-antiparticle pair, the amplitude of a certain production process from the initial state $\ket{I}$ is given in terms of the \textit{on-shell} $T$-matrix, $\bra{M\hat{p}\lambda\sigma}T\ket{I}$, where
\begin{equation}
    \ket{M\hat{p}\lambda\sigma}\equiv \ket{p_{+}p_{-}}\otimes\ket{\lambda\sigma}
\end{equation}
The first/second subspace in both momentum and spin corresponds to the particle/antiparticle, respectively, and $\lambda,\sigma$ label spin indices. Since energy is conserved from the initial state $I$, $M$ is fixed and the wave function describing the particle-antiparticle pair is
\begin{equation}
\ket{\Psi}=N\sum_{\lambda\sigma}\int\mathrm{d}\Omega~ \ket{M\hat{p}\lambda\sigma}\bra{M\hat{p}\lambda\sigma}T\ket{I}
\end{equation}
$\Omega$ being the solid angle related to $\hat{p}$ and $N$ some normalization factor. In high-energy experiments, typically only momentum measurements of the product particles are carried out. The resulting density matrix $\hat\varrho$ after a momentum POVM applied to $\ket{\Psi}$ is given in terms of the so-called production spin density matrix 
\begin{equation}\label{eq:ProductionSpinDensityMatrixT}
    R_{\lambda\sigma,\lambda'\sigma'}(M,\hat{p})\equiv \bra{M\hat{p}\lambda\sigma}T\ket{I}\bra{I}T^{\dagger}\ket{M\hat{p}\lambda'\sigma'}
\end{equation}
In this way,
\begin{equation}
\hat\varrho=\frac{1}{Z}\sum_{\lambda\sigma,\lambda'\sigma'}\int \mathrm{d}\Omega~R_{\lambda\sigma,\lambda'\sigma'}(M,\hat{p}) \frac{\ket{M\hat{p}\lambda\sigma}\bra{M\hat{p}\lambda'\sigma'}}{\braket{M\hat{p}|M\hat{p}}}
\end{equation}
with $Z=\int \mathrm{d}\Omega~\textrm{Tr}R(M,\hat{p})$ a normalization factor. The production spin density matrix is not properly normalized to unity, with its trace proportional to the differential cross-section of the particle-antiparticle production process at c.m. energy and direction $(M,\hat{p})$. A proper spin density matrix $\hat\varrho_{\lambda\sigma,\lambda'\sigma'}(M,\hat{p})$ is obtained as $\hat\varrho_{\lambda\sigma,\lambda'\sigma'}(M,\hat{p})=R_{\lambda\sigma,\lambda'\sigma'}(M,\hat{p})/\textrm{Tr}R$, which yields the following simple expression for the total quantum state 
\begin{equation}
\hat\varrho=\sum_{\lambda\sigma,\lambda'\sigma'}\int\mathrm{d}\Omega ~w(M,\hat{p})\hat\varrho_{\lambda\sigma,\lambda'\sigma'}(M,\hat{p}) \frac{\ket{M\hat{p}\lambda\sigma}\bra{M\hat{p}\lambda'\sigma'}}{\braket{M\hat{p}|M\hat{p}}}
\end{equation}
where the probability distribution $w(M,\hat{p})=\textrm{Tr}R(M,\hat{p})/Z$ is indeed normalized, $\int\mathrm{d}\Omega ~w(M,\hat{p})=1$. Thus, we can understand $\hat\varrho(M,\hat{p})$ as the density matrix describing the quantum state of the produced particle-antiparticle pair for fixed c.m. energy and momentum, and the total quantum state as the sum over all possible quantum states, weighted with the differential cross-section of each production process.

In general, as a $4\times 4$ Hermitian matrix, any production spin density matrix $R$ can be written in terms of direct products of the $2\times 2$ matrices $\sigma^\mu=[\sigma^0,\sigma^i]$, with $\sigma^i$ the usual Pauli matrices and $\sigma^0$ the identity. Specifically, $R$ is determined by $16$ parameters $\tilde{C}_{\mu\nu}$,
\begin{equation}\label{eq:ProductionSpinDensityMatrix}
R=\tilde{C}_{\mu\nu}\sigma^\mu\otimes\sigma^\nu
\end{equation}
with $\mathrm{Tr}R=4\tilde{C}_{00}$. The associated spin density matrix $\hat\varrho$, obtained by normalization, is then
\begin{equation}\label{eq:SpinDensityMatrix}
    \hat\varrho=\frac{C_{\mu\nu}\sigma^\mu\otimes\sigma^\nu}{4},~C_{\mu\nu}=\frac{\tilde{C}_{\mu\nu}}{\tilde{C}_{00}}
\end{equation}
By taking into account the trace orthogonality of the Pauli matrices, we have that $\textrm{tr}\left[\sigma^\mu\sigma^\nu\right]=2\delta^{\mu\nu}$, and thus we find that the $15$ coefficients
\begin{equation}\label{eq:SpinCorrelations}
    C_{\mu\nu}=\mathrm{tr}[\hat\varrho \sigma^{\mu}\otimes\sigma^{\nu}]
\end{equation}
provide precisely the expectation values of the spin observables. In particular, $C_{00}=1$ is fixed by normalization, $B_i=C_{i0},\bar{B}_i=C_{0i}$ are the vectors describing the particle/antiparticle spin polarizations, respectively, and the correlation matrix $C_{ij}$ describes the spin correlations between the particle and the antiparticle.

\section{Spin-momentum density matrix}
\label{sec:WaveFunction}
\subsection{Spin-momentum system in the rest frame}

In order to gain some insight on the relevant physics, and due to its simplicity and illustrative character, we choose for our study a simplified separable spin-momentum wave function for the particle/antiparticle (in the following labeled as systems $A,B$, respectively) with
antiparallel spins and momenta in the c.m. frame \cite{PhysRevA.81.042114}:
\begin{equation}
\ket{\psi}_{AB}
=\ket{\psi}_{p_Ap_B}\ket{\psi}_{\sigma_A\sigma_B},
\label{eq:Wavefunction_separable_form}
\end{equation}
where the spin and momentum states are defined as follows:
\begin{align}
&\ket{\psi}_{p_Ap_B} = \cos \alpha \ket{p_+p_-} + 
\sin\alpha\ket{p_-p_+},
\label{eq:Wavefunction_momentum}
\\
&\ket{\psi}_{\sigma_A\sigma_B} =\cos\beta\ket{\uparrow\downarrow} +
\sin\beta\ket{\downarrow\uparrow}.
\label{eq:WaveFunction_spin}
\end{align}
Here $\ket{\uparrow}$ and $\ket{\downarrow}$
denote $z$-projections of the spin operators.
We choose coordinate system with $z$-axis oriented along the momenta
of the particles, and therefore
$p_\pm = (p_0, 0, 0, \pm p_z)$.
In the rest frame, both wave functions have a similar mathematical structure 
and can be tackled similarly within a qubit formalism, as it is demonstrated 
in  Appendix \ref{sec:SimplestCase}. 
We note that the wave function Eq.(\ref{eq:Wavefunction_separable_form}) 
presents mirror symmetry under the transformation 
$A\leftrightarrow B$, 
$\alpha\rightarrow \pi/2-\alpha$, $\beta\rightarrow \pi/2-\beta$. 
Moreover, for $\alpha=\beta=\pi/4$, it also possess symmetry under spatial 
inversion, $\mathcal{P}\ket{\psi(\alpha)}_{p_Ap_B}=\ket{\psi(\alpha)}_{p_Ap_B}$, 
and discrete spin rotation 
$\mathcal{Z}_2\ket{\psi(\beta)}_{\sigma_A\sigma_B}=
\ket{\psi(\beta)}_{\sigma_A\sigma_B}$, $\mathcal{Z}_2=
e^{i\frac{\pi}{2}\sigma_x}$. 
Thus $\mathcal{P}\mathcal{Z}_2\ket{\psi(\alpha=\pi/4,\,
\beta=\pi/4)}_{AB}=\ket{\psi(\alpha=\pi/4,\,\beta=\pi/4)}_{AB}$. 
The same symmetries are generally expected in the boosted state. 
Before the boost, the full density matrix is the direct product of the 
momentum and spin matrices:
\begin{equation}
\hat\varrho_{AB} = 
\ket{\psi}_{AB}\bra{\psi}_{AB}
= \hat\varrho_{p_Ap_B} \otimes \hat\varrho_{\sigma_A\sigma_B}.
\label{eq:Full_density_matrix}
\end{equation}

For the two-particle system, we calculate the total entanglement for three different types of partitioning procedures applied to the four variables:
\begin{itemize}
\item $1+3$, the density matrix is reduced through the tracing of all possible combinations of $3$ degrees of freedom, and all contributions are summed;
\item $p+\sigma$, taking trace over the momentum and the spin separately;
\item $A+B$, bipartition into parts of particle $A$ and antiparticle $B$. Part $A$ contains the spin and the momentum of the particle, and part $B$ contains the spin and the momentum of the antiparticle.
\end{itemize}

Reducing the density matrix by all degrees of freedom 
except the momentum of the particle, we obtain
\begin{equation}
E(\hat\varrho_{p_A}) =\frac{ \sin^2 2\alpha} 2.
\label{eq:Alice_momentum_contributon_to_the_entanglement2}
\end{equation}
Similarly for the particle spin:
\begin{equation}
E(\hat \varrho_{\sigma_A}) =\frac{ \sin^2 2\beta } 2.
\label{eq:Alice_contributon_to_the_entanglement2}
\end{equation}
Taking into account similar contributions from the antiparticle and by summing
all contributions for the total entanglement we deduce:
\begin{equation}
E_{1+3}(\hat \varrho_{AB}) = \sin^2 2\alpha + \sin^2 2\beta.
\label{eq:Entropy_rho2}
\end{equation}

Partitioning into spin and momentum parts leads to zero entanglement
\begin{equation}
E_{p+\sigma}(\hat \varrho_{AB}) = 0.
\label{eq:Entanglement_spin_momentum}
\end{equation}
as $E(\hat\varrho_{p_Ap_B})=E(\hat\varrho_{\sigma_A\sigma_B})$. 

We proceed with partitioning $A+B$ when the remnant spin and momentum belong to the particle/antiparticle and the antiparticle/particle is traced out. The reduced matrix  after taking trace over the spin and momentum of the part $B$ reads:
\begin{align}
\hat\varrho_{p_A \sigma_A} = 
( \cos^2 \alpha \ket{p_+} & \bra{p_+}_A
  + \sin^2 \alpha \ket{p_-}\bra{p_-}_A
) 
\otimes
\nonumber
\\
& ( \cos^2 \beta \ket{\uparrow}\bra{\uparrow}_A
+ \sin^2 \beta \ket{\downarrow}\bra{\downarrow}_A). 
\label{eq:Density_matrix_A_spin_momentum}
\end{align}
Contribution of the state $\hat\varrho_{p_A \sigma_A}$ into the entanglement 
can be calculated straightforwardly and has the form:
\begin{align}
E(\hat \varrho_{p_A \sigma_A}) = &
1 - (\cos^4 \alpha + \sin^4 \alpha)(\cos^4 \beta + \sin^4 \beta) =
\nonumber
\\
& \frac{\sin^2 2\alpha} 2 + \frac{\sin^2 2 \beta} 2 -
\frac 1 4 \sin^2 2 \alpha \sin^2 2 \beta.
\label{eq:Alice_contributon_to_the_entanglement3}
\end{align}
The total entanglement then reads
\begin{equation}
E_{A+B}(\hat\varrho_{AB}) = \sin^2 2\alpha + \sin^2 2\beta - 
\frac 1 2 \sin^2 2\alpha \sin^2 2 \beta.
\label{eq:entanglement}
\end{equation}

\subsection{Measurement}
\label{subsec:Meausrement}

We perform a POVM along the $z$-projection of the spin of the particle.
According to Eq. (\ref{eq:DensityMatrix_AfterMeasurement}), the post-measurement 
density matrix has the form
\begin{equation}
\hat \varrho_{p_A \sigma_{Az}B} = \varrho_{{p_Ap_B}} \otimes
\left( \cos^2 \beta \ket{\uparrow\downarrow}\bra{\uparrow\downarrow} +
\sin^2  \beta \ket{\downarrow\uparrow}\bra{\downarrow\uparrow}  \right).
\label{eq:Measurement_spin_A}
\end{equation}
The expression obtained for $\hat \varrho_{p_A \sigma_{Az}B}$ 
is straightforwardly adapted to a momentum POVM through the replacement $\alpha \leftrightarrow \beta$ in the momentum sector.

Simultaneous measurement of the spin and momentum of particle $A$ 
yields the density matrix
\begin{eqnarray}
\hat \varrho_{p_{Az} \sigma_{Az} B} &=& 
(\cos^2 \alpha \ket{p_+p_-} \bra{p_+p_-} + 
\sin^2 \alpha \ket{p_-p_+}\bra{p_-p_+}
) \nonumber
\\
&\otimes& 
( \cos^2 \beta \ket{\uparrow\downarrow}\bra{\uparrow\downarrow}
+ \sin^2 \beta \ket{\downarrow\uparrow}\bra{\downarrow\uparrow}). 
\label{eq:Measurement_spin_momentum_A}
\end{eqnarray}
In the case of $1+3$ partitioning the entanglement in both cases is equal to:
\begin{eqnarray}
E_{1+3}(\hat \varrho_{p_{Az}\sigma_{Az}B}) &=& 
E_{1+3}(\hat \varrho_{p_{A}\sigma_{Az} B})  
= \sin^2 2\alpha + \sin^2 2\beta \nonumber\\ &=& E_{1+3}(\hat \varrho_{AB}). 
\label{eq:Entanglement_m_sA_one_plus_three}
\end{eqnarray}

On the other hand, partitioning into spin and momentum sectors 
leads to nonzero entanglement because of the non-vanishing
contribution of the spin part:
\begin{equation}
E_{p+\sigma}(\hat \varrho_{p_A\sigma_{Az}B}) = \frac {\sin^2 2 \beta} 2,
\label{eq:Entanglement_spin_A_two_plus_two}
\end{equation}
and
\begin{equation}
E_{p+\sigma}(\hat \varrho_{p_{Az}\sigma_{Az}B}) = \frac{\sin^2 2\alpha} 2 +
\frac{ \sin^2 2 \beta} 2.
\label{eq:Entanglement_spin_momentum_A_two_plus_two}
\end{equation}
Partitioning in two particles $A$ and $B$, i.e., 
$(\hat\sigma_A,\,p_A)$,  $(\hat\sigma_B,\,p_B)$
leads to the same result as in the previous
section:
\begin{eqnarray}\label{eq:Entanglement_measurement_spA_spB}
&&E_{A+B}(\hat \varrho_{p_A\sigma_{Az}B}) = 
E_{A+B} (\hat \varrho_{p_{Az}\sigma_{Az}B})=\nonumber\\
&&\sin^2 2\alpha + \sin^2 2\beta - 
\frac 1 2 \sin^2 2\alpha \sin^2 2 \beta.
\end{eqnarray}
Same results would have been obtained for POVMs on the antiparticle $B$.

\subsection{Conditional entropies}
\label{subsec:Conditional}

The conditional entropy of a subsystem is found from Eq. (\ref{eq:Conditional_entropy_definition}).
Let $AB$ be an entire system depending on the spin and momenta of both 
$A$ and $B$ subsystems. 
The entropy of the entire system depends on the type of partition chosen. 
The entropy of a particular subsystem is defined
after tracing of the second subsystem.
The entropy of the spin part of $A$ was found in
Eq. (\ref{eq:Alice_contributon_to_the_entanglement2}).
The relative entropy of the spin part of $A$ follows the same type 
of partitioning $1+3$ and is equal to
\begin{eqnarray}
&&E_{1+3}(\{p_A,p_B,\sigma_B\}|\sigma_A) =
E_{1+3}(\hat\varrho_{AB}) - E(\hat\varrho_{\sigma_A}) =\nonumber\\ 
&&\sin^2 2\alpha +\frac {\sin^2 2\beta} 2.
\label{eq:re1plus3}
\end{eqnarray}
For $p+\sigma$ partition we have:
\begin{eqnarray}
E_{p+\sigma}(\{p_A,p_B,\sigma_B\}|\sigma_A) = -\frac{\sin^2 2\beta} 2,
\label{eq:Relative_entropy_spin_A_ppluss}
\end{eqnarray}
and for partition $A+B$ we obtain:
\begin{eqnarray}
&&E_{A+B}(\{p_A,p_B,\sigma_B\}|\sigma_A) = 
\sin^2 2\alpha + \frac{\sin^2 2 \beta}{2}-\nonumber\\
&&\frac {1}{2}\sin^2 2\alpha \sin^2 2\beta.
\label{eq:Relative_entropy_spin_A_A+B}
\end{eqnarray}
The entropy of the subsystems of two spins and two momenta
is equal to zero in the initial state:
\begin{equation}
E(\hat\varrho_{\sigma_A\sigma_B}) = E(\hat\varrho_{p_Ap_B}) =0.
\label{eq:Entropy_zero}
\end{equation}
We calculate the relative entropy for $1+3$ partition:
\begin{eqnarray}
E_{1+3}(\{p_A,p_B\}\vert\{\sigma_A,\sigma_B\}) &=& E_{1+3}(\hat \varrho_{AB})-E(\hat \varrho_{\sigma_A\sigma_B})
\nonumber
\\
&=&
\sin^2 2\alpha+\sin^2 2\beta,
\label{eq:Entropy_ppss_1plus3}
\end{eqnarray}
as well as for the remaining partitions
\begin{eqnarray}
E_{p+\sigma}(\{p_A,p_B\}\vert\{\sigma_A,\sigma_B\}) &=&0,
\\
E_{A+B}(\{p_A,p_B\}\vert\{\sigma_A,\sigma_B\}) &=&\sin^2 2\alpha + \sin^2 2\beta
\label{eq:Entropy_ppss_ppluss}
\nonumber
\\
&-&\frac 1 2\sin^2 2\alpha \sin^2 2\beta.
\label{eq:Entropy_ppss_AplusB}
\end{eqnarray}
Similar relations hold for relative entropies of the spins and the momenta.

We also can calculate the relative entropy between particles $A$ and $B$.
Using Eq. (\ref{eq:Density_matrix_A_spin_momentum}) we easily obtain:

\begin{eqnarray}
&&E_{1+3}(\{p_B, \sigma_B\}\vert \{p_A,\sigma_A\}) =
\frac{\sin^2 2\alpha} 2 + \frac {\sin^2 2\beta} 2 +\nonumber\\ 
&&\frac 1 4 \sin^2 2\alpha \sin^2 2\beta, 
\label{eq:Relative_entropy_spinmomentumA_1plus3}
\end{eqnarray}
\begin{eqnarray}
&&E_{p+\sigma}(\{p_B, \sigma_B\}\vert \{p_A,\sigma_A\}) = 
-\frac{\sin^2 2\alpha} 2 - \frac {\sin^2 2\beta} 2 +\nonumber\\ 
&&\frac 1 4 \sin^2 2\alpha \sin^2 2\beta, 
\label{eq:relative_entropy_spinmomentumA_ppluss}
\end{eqnarray}
\begin{eqnarray}
&&E_{A+B}(\{p_B, \sigma_B\}\vert \{p_A,\sigma_A\})  =
\frac{\sin^2 2\alpha} 2 + \frac {\sin^2 2\beta} 2 -\nonumber\\ 
&&\frac 1 4 \sin^2 2\alpha \sin^2 2\beta. 
\label{eq:relative_entropy_spinmomentuma_AplusB}
\end{eqnarray}
The mutual information between the spins and momenta is then determined by the formulae:
\begin{align}
I(p_A, p_B) & = \sin^2 2\alpha,
\label{eq:MutualInformation_momentum_A_momentum_B}
\\
I(\sigma_A, \sigma_B) & = \sin^2 2\beta.
\label{eq:Mutual_information_spin_A_spin_B}
\\
I(\sigma_A, p_A) & = \frac 1 4 \sin^2 2\alpha \sin^2 2\beta.
\label{eq:Mutual_information_spin_A_momentum_A}
\end{align}
The mutual information between the subsystems of spins and momenta is zero
\begin{equation}
I(\{\sigma_A,\sigma_b\}, \{p_A,p_B\}) = 0. 
\label{Mutual_information_spin_AB_momentum_AB}
\end{equation}

\section{Relativistic transformations}
\label{sec:Relativistic}
\subsection{Influence of the Wigner rotations on entanglement}

The Lorentz group contains boosts, which are rotation-free Lorentz transformations connecting two uniformly moving frames,
plus rotations. We follow the same notation of Weinberg \cite{weinberg1995quantum}, and introduce Lorentz transformations 
$\hat T(\Lambda)$ of the coordinates, $x'^\mu=\Lambda^\mu_\nu x^\nu$. From the generators of boosts $\hat K$ and rotations $\hat J$, one can construct two new generators 
$\hat A=1/2(\hat J+i\hat K)$ and $\hat B=1/2(\hat J-i\hat K)$. These new generators form closed algebras and therefore are equivalent to the direct product of two groups $SU(2)\otimes SU(2)$. Due to this fact, the formal action of two subsequent boosts is equivalent to the action of a boost and a rotation.


At the quantum level, we consider the effect of a Lorentz transformation $\hat T(\Lambda)$ on the eigenstate of the four-momentum operator $\hat P^\mu$:
\begin{eqnarray}\label{four momentum}
\hat P^\mu \ket{p,\sigma}=p^\mu\ket{p,\sigma}.
\end{eqnarray}
The effect of the coordinate transformation $\hat T(\Lambda)$ on any state $\ket{\psi}$ is described by the operator $\hat U(\Lambda)$, $\ket{\psi'}=\hat U(\Lambda) \ket{\psi}$. Taking into account that
\begin{eqnarray}\label{Lorentz group transformation}
\hat U(\Lambda)\hat P^\gamma\hat U^{-1}(\Lambda)=\Lambda_\mu^\gamma \hat P^\mu,
\end{eqnarray}
and, by comparing with Eq.(\ref{four momentum}), we infer that $\hat U(\Lambda)\ket{p,\sigma}$ must be a linear combination of the boosted momentum  eigenstates:
\begin{eqnarray}\label{Symmetry combination}
\hat U(\Lambda)\ket{p,\sigma}=\sum\limits_{\sigma'} W_{\sigma'\sigma}(\Lambda,p)\ket{\Lambda p,\sigma'}\equiv \ket{\Lambda p, \sigma_\Lambda}.
\end{eqnarray}
Thus, Lorentz transformations act on both momentum and spin variables. We can compute explicitly the matrix $W_{\sigma'\sigma}(\Lambda,p)$ by noting that, if the particle is massive, any state $\ket{p,\sigma}$ can be in turn obtained from the rest frame state $\ket{k,\sigma}$ through the appropiate Lorentz transformation $L[p]$:
\begin{equation}
\ket{p,\sigma} = \hat U(L[p]) \ket{k,\sigma}.
\label{eq:Lorentz_transformFromRestFrame}
\end{equation}
where we stress that a boost from the rest frame does not modify the spin $\sigma$.

Thus, concatenating both Lorentz transformations gives
\begin{eqnarray}
\ket{\Lambda p, \sigma_\Lambda}&=&\hat U(\Lambda) \ket{p,\sigma} = \hat U(\Lambda L[p]) \ket{k,\sigma} \label{eq:Lorentz_transformFromRestFrame_Total}\\
\nonumber &=&\hat U(L[\Lambda p])\hat U(L^{-1}[\Lambda p]\Lambda L[p]) \ket{k,\sigma}
\end{eqnarray}
We note that the total Lorentz transformation $\hat U(L^{-1}[\Lambda p]\Lambda L[p])$ boosts twice, and then goes back to the rest frame. Therefore, from Eq. (\ref{Symmetry combination}), we conclude that its effect amounts to a spin rotation, known as the \textit{Wigner rotation}, described precisely by the matrix $W_{\sigma'\sigma}(\Lambda,p)$:
\begin{equation}
    \ket{\Lambda p, \sigma_\Lambda}=\sum\limits_{\sigma'} W_{\sigma'\sigma}(\Lambda,p)\ket{\Lambda p, \sigma'}
\end{equation}
In the following, we consider that the initial momentum $p$ is aligned along 
the $z$-axis, while the boost $\Lambda$ is perpendicular and chosen along 
the $x$-axis without loss of generality. 
After these two consecutive boosts \cite{alsing2002lorentz}, the Wigner 
rotation matrix $W(\Lambda,p)$ is characterized by the parameter 
\begin{equation}
\tan \delta = \frac{\sinh \xi \sinh \eta}
{\cosh \xi + \cosh \eta},
\label{eq:Wigner_rotation}
\end{equation}
where $\xi$ and $\eta$ are the hyperbolic rotation angles corresponding
to the boosts along the $z$ and $x$ axes, respectively: 
$\tanh\xi=v_z/c$, $\tanh\eta=v_x/c$.
\begin{figure}[t!]
\centering
\includegraphics[width=\columnwidth]{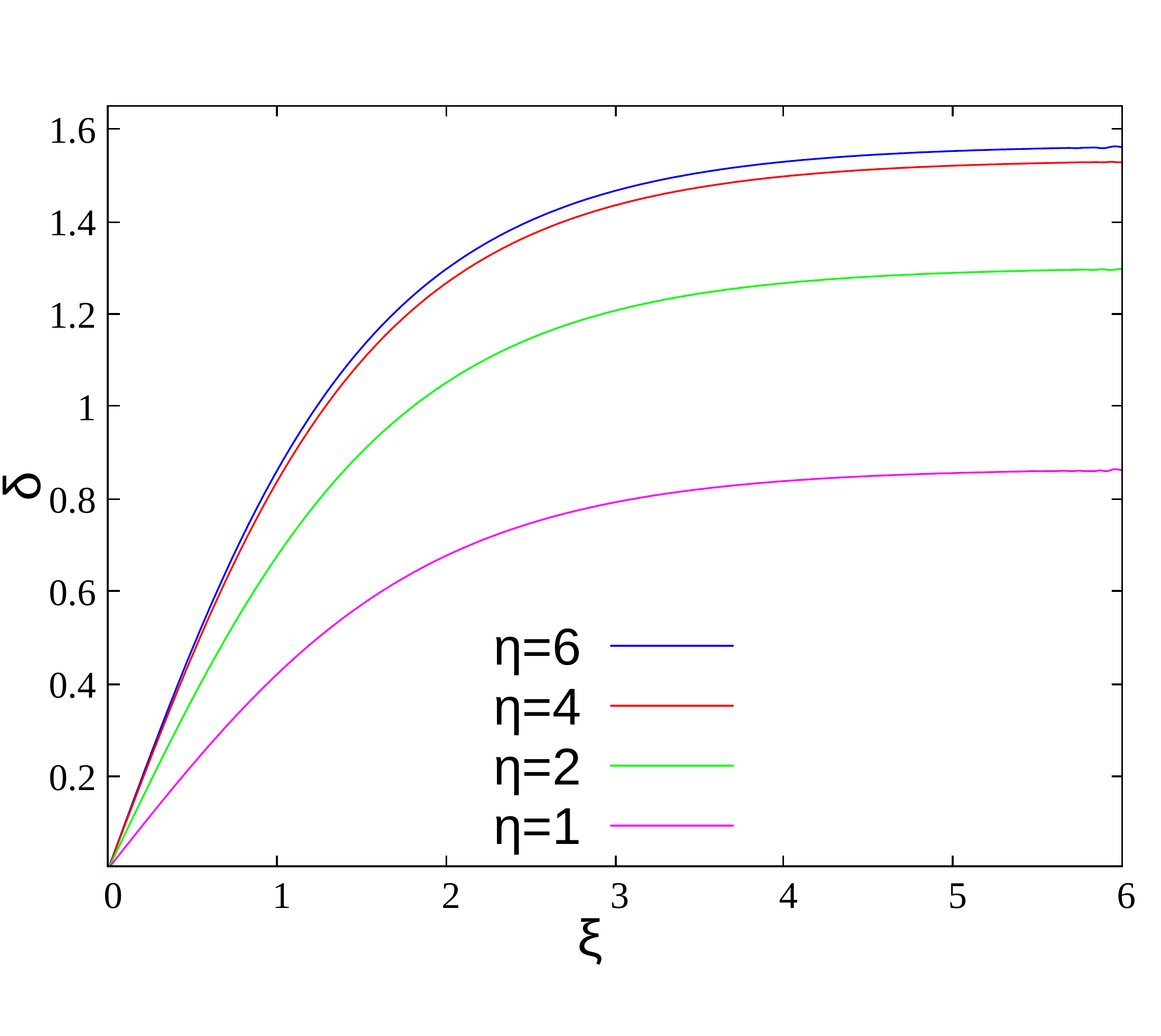}
\caption{Dependence of the Wigner rotation angle $\delta$
on the parameter $\xi$ for different values of $\eta$,
plotted through Eq. (\ref{eq:Wigner_rotation}).  The parameter $\xi$ 
characterizes the boost along $x$-direction, while $\eta$ is the boost parameter
of the particle moving in the rest frame.}
\label{fig:Hyperbolic}
\end{figure}
Figure \ref{fig:Hyperbolic} represents the dependence of the
Wigner angle $\delta$ on $\xi$ and $\eta$. 
One can see that, for $\xi,\eta \gg 1$, the Wigner angle $\delta$ approaches $\pi/2$.

We now apply this formalism to the quantum state of Eq. (\ref{eq:Wavefunction_separable_form}). In particular, we consider a boost in the c.m. frame perpendicular to the anti-parallel particle momenta $p_{\pm}$. Consequently, the total boosted state can be written as
\begin{eqnarray}
&&\ket{\Psi_\Lambda}_{\mathrm{total}} = 
\cos \alpha \ket{\Lambda p_+, \Lambda p_-} (U_+ \otimes U_-)
\ket{\psi}_{\mathrm{spin}}+\nonumber\\
&&\sin \alpha \ket{\Lambda p_-, \Lambda p_+} (U_- \otimes U_+)
\ket{\psi}_{\mathrm{spin}},
\label{eq:State_total_boosted}
\end{eqnarray}
where $U_\pm$ are the spin rotation matrices describing the transformation
from the rest frame of a particle with momentum $p_{\pm}$ to the final boosted state:
\begin{equation}
U_\pm = 
\begin{pmatrix}
\cos \frac \delta 2 && \pm \sin \frac \delta 2 \\
\mp \sin \frac \delta 2 && \cos \frac \delta 2
\end{pmatrix}.
\label{eq:SpinU}
\end{equation}
By expanding these transformations, we get
\begin{equation}
\ket{\Psi_\Lambda}_{\mathrm{total}} = 
\cos \alpha \ket{\Lambda p_+,\Lambda p_-}\ket{a_1} +
\sin \alpha \ket{\Lambda p_-,\Lambda p_+}\ket{a_2},
\label{eq:Psi_total_lambda}
\end{equation}
where 
\begin{eqnarray}
\ket{a_1}&\equiv&c_1 \ket {\uparrow \downarrow} +c_2 \ket{\downarrow \uparrow} +
c_{3}\ket{\uparrow\uparrow} + c_4\ket{\downarrow \downarrow}\\
\ket{a_2}&\equiv&c_1 \ket {\uparrow \downarrow} +c_2 \ket{\downarrow \uparrow} -
c_{3}\ket{\uparrow\uparrow} - c_4\ket{\downarrow \downarrow}
\end{eqnarray}
and 
\begin{align}
c_1  & = \frac 1 2 (\cos \beta + \sin \beta) + 
\frac {\cos \delta} 2 (\cos \beta -\sin \beta),
\label{eq:c1}
\\
c_2 & = \frac 1 2 (\cos \beta + \sin \beta) -
\frac {\cos\delta} 2 (\cos \beta -\sin \beta),
\label{eq:c2}
\\
c_3 & = \frac 1 2 \sin \delta (\sin \beta - \cos \beta),
\label{eq:c3}
\\
c_4 & =c_3.
\label{eq:c4}
\end{align}
The total density matrix of the boosted system then reads
\begin{widetext}
\begin{eqnarray}
\nonumber \hat \varrho_{AB}^\Lambda &=& 
\cos^2 \alpha 
\ket{\Lambda p_+,\Lambda p_-}\bra{\Lambda p_+ \Lambda p_-} \otimes A_{11}+\sin \alpha \cos \alpha\left( 
\ket{\Lambda p_+ \Lambda p_-}\bra{\Lambda p_- \Lambda p_+}\otimes A_{12}  +
\ket{\Lambda p_- \Lambda p_+}\bra{\Lambda p_+ \Lambda p_-} \otimes A_{21}
\right) 
\\ 
&+&\sin^2\alpha
\ket{\Lambda p_- \Lambda p_+}\bra{\Lambda p_- \Lambda p_+}\otimes A_{22},\label{eq:FullRhoLambda} 
\end{eqnarray}
\end{widetext}
where the $4\times4$ spin matrices $A_{ik},~i,k=1,2$, result from the direct products of 
the spin states $\ket{a_i}$ at the r.h.s. of 
Eq. (\ref{eq:Psi_total_lambda}):
\begin{equation}
A_{ik} = \ket{a_i} \otimes \bra{a_k},
\label{eq:Aik_text}
\end{equation}
We refer the reader to Appendix \ref{sec:BoostedMatrices}
for the derivation of useful relations between the coefficients $c_i$ and the
properties of the spin density matrices $A_{ik}$. The contribution of momentum degrees of freedom is similar to the
unboosted case:
\begin{equation}
E(\hat\varrho^{\Lambda}_{p_A}) = E(\hat\varrho^{\Lambda}_{p_B})=\frac{\sin^2 2\alpha}{2}.
\label{eq:Entanglement_momentum_boosted}
\end{equation}
Due to the Wigner rotation, the final expressions for spin degrees of freedom are more involved. To derive them, we trace the momentum degrees of freedom.
The resulting spin density matrices are expressed in
Eqs. (\ref{eq:DensityMatrix_spins_boosted}) and 
(\ref{eq:Subsystem_of_Alice_spin_boosted}) in
Appendix \ref{sec:BoostedMatrices}.
Using Eqs. (\ref{eq:c1})-(\ref{eq:c4}),
one deduces explicitly the spin contribution:
\begin{equation}
E(\hat \varrho^{\Lambda}_{\sigma_A}) = E(\hat \varrho^{\Lambda}_{\sigma_B}) =\frac {\sin^2 2\beta} 2
+ \frac 1 2 \sin^2 \delta \sin^2 2 \alpha \cos^2 2 \beta.
\label{eq:Alice_spin_contributon_to_the_entanglement_boosted}
\end{equation}
After adding both the spin and momentum contribution, we finally obtain
\begin{equation}
E_{1+3}(\hat \varrho^{\Lambda}_{AB}) = 
\sin^2 2\alpha + \sin^2 2 \beta +
\sin^2 \delta \sin^2 2\alpha \cos^2 2 \beta .
\label{eq:Entropy_rho_boosted}
\end{equation}

Partitioning into spin and momentum degrees of freedom leads to zero entanglement in the rest frame. In the case of the boosted system, one can show that the
sums of squares of matrix elements in Eq. 
(\ref{eq:DensityMatrix_spins_boosted}) and (\ref{eq:DensityMatrix_momenta_boosted})
are identical, $E(\hat \varrho^\Lambda_{\sigma_A \sigma_B})=E(\hat \varrho^\Lambda_{p_A p_B})$, with 
\begin{equation}
    E(\hat \varrho^\Lambda_{\sigma_A \sigma_B}) =
\frac 1 2 \sin^2 \delta \sin^2 2\alpha 
[\cos^2 2\beta + \cos^2 \delta (1-\sin 2\beta)^2].
\label{eq:Entropy_sAsB_boosted_new}
\end{equation}
and thus
\begin{equation}
E_{p+\sigma}(\hat\varrho^\Lambda_{AB}) = 
\sin^2 \delta \sin^2 2\alpha 
\left[ \cos^2 2\beta + \cos^2 \delta (1-\sin 2\beta)^2\right].
\label{eq:Entanglement_spin_plus_momentum_boosted}
\end{equation}

Finally, for the $A+B$ partition, we get
\begin{equation}
E_{A+B}(\varrho^\Lambda_{AB}) = \sin^2 2\alpha + \sin^2 2\beta -\frac 1 2
\sin^2 2 \alpha \sin^2 2\beta.
\label{eq:Entanglement_A_spin_momentum_boosted}
\end{equation}
The obtained result $E_{A+B}(\varrho^\Lambda)$ does not depend on the boost parameter $\delta$ and
coincides with Eq. (\ref{eq:entanglement}). This results from the relativistic invariance of the entropy of each subsystem $A$ and $B$.

\subsection{Measurement}
\label{subsec:measurement2}

We perform a POVM on the boosted state and explore the entanglement of the resulting post-measurement  state.
At first, we consider spin measurements. 
The post-measurement matrix after measuring the spin of $A$ is given by 
$\hat \varrho^{\Lambda}_{\sigma_A}$, whose explicit expression is provided in Appendix \ref{sec:DerivationMeasurement}.
To find the spin contribution in the $1+3$ partition, we trace over the momentum variables. The contribution to the entropy is equal to
\begin{eqnarray}
&&E(\hat \rho^\Lambda_{\sigma_{Az}}) = 
\frac 1 2 \sin^2 2\beta +\frac 1 2 \sin^2 \delta \cos^2 2\beta. 
\label{eq:Alice_spin_contributon_to_the_entanglement_measured_boosted}
\end{eqnarray}
The total entanglement is shown to be
\begin{equation}
E_{1+3}(\hat \rho^\Lambda_{p_A\sigma_{Az}B}) = 
\sin^2 2\alpha +\sin^2 2\beta +\sin^2 \delta \cos^2 2\beta.
\label{eq:Entanglement_full_sA_measured_boosted}
\end{equation}
This result is different from Eq. (\ref{eq:Entropy_rho_boosted}) since the last
term on the r.h.s. does not depend on $\alpha$. For the spin-momentum partition we have:
\begin{widetext}
\begin{eqnarray}
&&E_{p+\sigma}(\hat \rho_{p_A\sigma_{Az}B}^\Lambda)= \frac 3 4
\sin^2 \delta \sin^2 2 \alpha\left[\cos^2 2\beta + \cos^2 \delta (1-\sin 2\beta)^2 \right]+\frac{ \sin^2 2 \beta }2  + \frac 1 2 \sin^2 \delta \cos^2 2\beta.
\label{eq:Entropy_SpinMomentumSplit_sA_measured_boosted}
\end{eqnarray}
In the limit $\delta \rightarrow 0$ we recover
Eq. (\ref{eq:Entanglement_spin_A_two_plus_two}).
Similarly 
\begin{equation}
E_{p+\sigma}(\hat \varrho^\Lambda_{p_{Az}\sigma_{Az}B})= \frac 1 4
\sin^2 \delta \sin^2 2 \alpha \left[\cos^2 2\beta + \cos^2 \delta 
(1-\sin 2\beta)^2 \right] +\frac {\sin^2 2 \alpha} 2 + \frac{ \sin^2 2 \beta }2  
+ \frac 1 2 \sin^2 \delta \cos^2 2\beta,
\label{eq:Entropy_SpinMomentumSplit_sApA_measured_boosted}
\end{equation}
\end{widetext}
converts into Eq. (\ref{eq:Entanglement_spin_momentum_A_two_plus_two})
as $\delta \rightarrow 0$. After partitioning into each subsystem $A$ and $B$ one obtains:\\
\begin{eqnarray}
&& E_{A+B}(\hat \varrho^\Lambda_{p_A\sigma_{Az}B}) = 
E_{A+B} (\hat \varrho^\Lambda_{p_{Az}\sigma_{Az}B}) = 
\sin^2 2\alpha + \sin^2 2\beta -\nonumber\\ 
&&\frac 1 2 \sin^2 2\alpha \sin^2 2 \beta +\frac 1 2\sin^2 \delta \cos^2 2\beta  \left(1 - \frac {\sin^2 2\alpha} 2\right). 
\label{eq:Entanglement_measurement_spA_spB_boosted}
\end{eqnarray}
Remarkably, unlike in Eqs. (\ref{eq:entanglement}), (\ref{eq:Entanglement_measurement_spA_spB})
and (\ref{eq:Entanglement_A_spin_momentum_boosted}),
now there is an extra term in Eq.(\ref{eq:Entanglement_measurement_spA_spB_boosted}) 
which does depend on the strength of the boost $\delta$.

\subsection{Conditional entropies}
\label{subsec:Conditional2}

Let us consider conditional entropies and mutual information of the boosted system. In analogy with
Subsection \ref{subsec:Conditional}, the relative entropy with respect to the spin $A$ in the case of $1+3$ partition reads:
\begin{eqnarray}
&&E^\Lambda_{1+3}(\{p_A, p_B, \sigma_B\}|\sigma_A) = 
\sin^2 2 \alpha + \frac {\sin^2 2 \beta} 2 +\nonumber\\
&&\frac 1 2 \sin^2 \delta \sin^2 2 \alpha \cos^2 2\beta.
\label{eq:Relative_entropy_spin_A_1plus3_boosted}
\end{eqnarray}
For the spin-momentum and $A+B$ partitions we deduce:
\begin{align}
&E^\Lambda_{p+\sigma}(\{p_A, p_B, \sigma_B\}|\sigma_A)= 
-\frac{\sin^2 2\beta} 2\label{eq:Relative_entropy_spin_A_ppluss_boosted}\\
&+\sin^2 \delta \sin^2 2\alpha  \left[\frac {\cos^2 2\beta} 2 + 
\cos^2 \delta (1- \sin 2\beta)^2 \right],\nonumber
\\
&E^\Lambda_{A+B}(\{p_A, p_B, \sigma_B\}|\sigma_A) =   
\sin^2 2\alpha + \frac {\sin^2 2\beta} 2 \label{eq:Relative_entropy_spin_A_B_boosted}\\
&-\frac 1 2 \sin^2 2\alpha \sin^2 2\beta -\frac 1 2 \sin^2 \delta \sin^2 2\alpha \cos^2 2\beta.\nonumber
\end{align}

As we already have seen, the entropy of the spin and momentum subsystems is not zero after the boost.
Therefore, the relative entropies between the spins and momenta are expressed by following formulae:
\begin{widetext}

\begin{eqnarray}
E^\Lambda_{1+3}(\{p_A, p_B\}|\{\sigma_A, \sigma_B\})
&=&\sin^2 2\alpha +\sin^2 2\beta+\frac 1 2 \sin^2 \delta \sin^2 2 \alpha \left[ \cos^2 2\beta + 
\cos^2 \delta (1 - \sin 2 \beta)^2 \right],
\label{eq:Relative_entropy_pp_ss_1plus3_boosted}
\\
E^\Lambda_{p+\sigma}(\{p_A, p_B\}|\{\sigma_A, \sigma_B\}) &=& 
\frac 1 2 \sin^2 \delta \sin^2 2\alpha \left[ \cos^2 2\beta - \cos^2 \delta (1 - \sin 2 \beta)^2 \right],
\label{eq:Relative_entropy_pp_ss_1plus3_boosted}\\
E^\Lambda_{A+B}(\{p_A, p_B\}|\{\sigma_A, \sigma_B\})&=& 
\sin^2 2\alpha + \sin^2 2\beta -\frac 1 2 \cos^2 \delta \sin^2 2\alpha \sin^2 2\beta 
-\sin^2 \delta \sin^2 2\alpha [1 + \cos^2 \delta (1-\sin 2\beta)^2].
\label{eq:Relative_entropy_AB_1plus3_boosted}
\end{eqnarray}
The relative entropies between the particles $A$ and $B$ are equal to
\begin{eqnarray}
E^\Lambda_{1+3}(\{p_B, \sigma_B\}|\{p_A, \sigma_A\})&=&
\frac {\sin^2 2\alpha} 2 + \frac {\sin^2 2\beta} 2 +\frac 1 4 \sin^2 2\alpha \sin^2 2\beta +\sin^2 \delta \sin^2 2\alpha \cos^2 2\beta,
\label{eq:Relative_entropy_spinmomentumA_1plus3_boosted}\\
E^\Lambda_{p+\sigma}(\{p_B, \sigma_B\}|\{p_A, \sigma_A\})&=&-\frac {\sin^2 2\alpha} 2 - \frac {\sin^2 2\beta} 2 +\frac 1 4 \sin^2 2\alpha \sin^2 2\beta\nonumber\\&+&
\sin^2\delta \sin^2 2\alpha[\cos^2 2\beta + \cos^2\delta (1-\sin 2\beta)^2] ,
\label{eq:relative_entropy_spinmomentumA_ppluss_boosted}
\\
E^\Lambda_{A+B}(\{p_B, \sigma_B\}|\{p_A, \sigma_A\}) &=& 
\frac {\sin^2 2\alpha} 2 + \frac{\sin^2 2\beta} 2 -\frac 1 4 \sin^2 2\alpha \sin^2 2 \beta.
\label{eq:relative_entropy_spinmomentuma_AplusB_boosted}
\end{eqnarray}

The mutual information between the spins and the momenta takes the form:
\begin{eqnarray}
I_\Lambda(p_A, p_B) & =& \sin^2 2\alpha - \frac 1 2
\sin^2 \delta \sin^2 2\alpha 
[\cos^2 2\beta + \cos^2 \delta (1 - \sin^2 2 \beta)^2],
\label{eq:MutualInformation_momentum_A_momentum_B_boosted}\\
I_\Lambda(\sigma_A, \sigma_B) & =& \sin^2 2\beta + \frac 1 2
\sin^2 \delta \sin^2 2 \alpha
[\cos^2 2\beta -\cos^2 \delta (1-\sin 2\beta)^2],
\label{eq:Mutual_information_spin_A_spin_B_boosted}
\\
I_\Lambda(\sigma_A, p_A) &=& \frac 1 4 \sin^2 2\alpha  \sin^2 2\beta
+ \frac 1 2 \sin^2 \delta \sin^2 2\alpha \cos^2 2\beta.
\label{eq:Mutual_information_spin_A_momentum_A_bosted}
\end{eqnarray}
Finally, the mutual information between the subsystems of spins and momenta reads:
\begin{equation}
I_\Lambda(\{\sigma_A,\sigma_B\}, \{p_A,p_B\}) = 
\sin^2 \delta \sin^2 2 \alpha
[\cos^2 2\beta + \cos^2 \delta (1-\sin 2\beta)^2].
\label{Mutual_information_spin_AB_momentum_AB_boosted}
\end{equation}
\end{widetext}
\section{Quantum discord} 
\label{sec:Discord}
We explore the quantum discord of the particle-antiparticle pair 
(denoted as Alice $A$ and Bob $B$, respectively) for arbitrary boost. 
We also analyze the quantum discord of the boosted 
post-measurement state after performing a 
POVM on the spin of one of the particles of the pair. 

\subsection{General expressions and definitions}
\label{subsec:definitions}

The quantum discord is expressed by
\begin{equation}
\label{eq:QuantumDiscordGeneralExpression}
D =\min_{\{\hat \Pi_j^B\}}\{E(B)-E(A,B)+E(A\vert \{\hat\Pi_j^B\})\},
\end{equation}
where $E(A)$ and $E(A,B)$ are calculated from their respective density
matrices, $\hat\varrho_A = \mathrm{Tr}_B (\hat \varrho_{AB})$ and
$\hat \varrho_{AB}$, according to the expression for the entropy:
\begin{equation}
\label{eq:EntropyGeneralExpression}
E(\hat \varrho) = \sum_i (1- \mathrm{Tr}(\varrho_i^2)) =
\sum_i (1 - \vert \rho_{i,\, mn} \vert^2 ).
\end{equation}
Projected states of the density matrix:
\begin{equation}
\label{eq:ProjectedStates}
\hat\varrho_{A\vert \hat \Pi_j^B} =
\hat \Pi_j^B \hat \varrho_{AB} \hat \Pi_j^B/p_j,
\end{equation}
where the probabilities $p_j$ are expressed as
\begin{equation}
\label{eq:ProjectionProbabilities}
p_j = \mathrm{Tr}(\hat \varrho_{AB} \hat \Pi_j^B).
\end{equation}
The combined entropy of the projected states is equal to
\begin{equation}
\label{eq:ProjectedEntropyDefinition}
E(A\vert \{\hat\Pi_j^B\}) = 
\sum_j p_j E(\hat \varrho_{A\vert \hat \Pi_j^B}).
\end{equation}
Define quantum states corresponding to opposite orientations
of spins along any arbitrary direction $s$
defined by all possible states of the spin of particle $B$
\cite{PhysRevB.78.224413}:
\begin{eqnarray}
&&\ket{s_-}_B = \cos \theta/2 \ket{\downarrow}_B + \sin \theta/2 e^{i\varphi}
\ket{\uparrow}_B,
\label{eq:Sminus}
\\
&&\ket{s_+}_B= e^{-i\varphi}\sin \theta/2 \ket{\downarrow}_B -
\cos \theta/2 \ket{\uparrow}_B,
\label{eq:Splus}
\end{eqnarray}
where the angle parameters $\theta$ and $\varphi$ span the Bloch sphere:
\begin{equation}
\label{eq:Theta_Phi_limit}
0 \leq \theta \leq \pi, \, 0 \leq \varphi < 2\pi.
\end{equation}
For the case when the two variables are the  spins $\sigma_A$ and $\sigma_B$, 
the first and the second terms 
in Eq. (\ref{eq:QuantumDiscordGeneralExpression}), according to the previous
results, are expressed as
\begin{eqnarray}
&& E(\hat \varrho^\Lambda_{\sigma_B}) = \frac {\sin^2 2\beta} 2
+ \frac 1 2 \sin^2 \delta \sin^2 2\alpha \cos^2 2\beta,
\label{eq:EntropyOneSpin}
\\
&& E(\hat\varrho^\Lambda_{\sigma_A\sigma_B}) = \frac 1 2
\sin^2 \delta \sin^2 2\alpha [\cos^2 2\beta + 
\nonumber
\\
&& \qquad  \qquad \qquad \cos^2\delta
(1-\sin 2 \beta)^2],
\label{eq:EntropyTwoSpins}
\end{eqnarray}
and their combination yields
\begin{eqnarray}
&& E(\varrho^\Lambda_{\sigma_B}) - E(\hat\varrho^\Lambda_{\sigma_A\sigma_B}) =
\frac {\sin^2 2\beta}  2 -
\nonumber
\\
&& \qquad \qquad
\frac 1 2 \sin^2 \delta \cos^2\delta{\sin^2 2\alpha} (1-\sin 2\beta)^2.
\label{eq:CombinationOfEntropies}
\end{eqnarray}

Using Eqs. (\ref{eq:Bup}) and (\ref{eq:Bdown}) 
from Appendix B, it is possible to  write down
spin projection operators correponding to $s_\pm$ in
Eqs. (\ref{eq:Sminus}) and (\ref{eq:Splus}):
\begin{widetext}
\begin{equation}
\label{eq:PiPlusMinus}
\hat \Pi^{\sigma_B}_{s_\pm} = \frac 1 2
\begin{pmatrix}
1\mp\cos\theta && 0 && \mp e^{i\varphi} \sin \theta && 0 \\
0 && 1\pm\cos \theta && 0 && \mp e^{-i\varphi} \sin \theta \\
\mp e^{-i\varphi}\sin \theta && 0 && 1\pm \cos \theta && 0 \\
0 && \mp e^{i\varphi} \sin \theta && 0 && 1\mp\cos \theta
\end{pmatrix}.
\end{equation}
Setting notation 
$t = \tan \theta/2$,
\begin{equation}
\hat \Pi^{\sigma_B}_{s_+} = \frac{1}{1+t^2}
\begin{pmatrix}
t^2 && 0 && -t e^{i\varphi}  && 0 \\
0 && 1 && 0 && -te^{-i\varphi} \\
-t e^{-i\varphi} && 0 && 1 && 0 \\
0 && -t e^{i\varphi} && 0 && t^2 
\end{pmatrix}
\label{eq:PiPlus_t}
\end{equation}
and
\begin{equation}
\hat \Pi^{\sigma_B}_{s_-} = \frac{1}{1+t^2}
\begin{pmatrix}
1 && 0 && te^{i\varphi} && 0 \\
0 && t^2&& 0 && te^{-i\varphi} \\
te^{-i\varphi}&& 0 && t^2 && 0 \\
0 && t e^{i\varphi} && 0 && 1 
\end{pmatrix}
\label{eq:PiMinus_t}
\end{equation}
\end{widetext}
\subsection{Quantum discord for arbitrary boost}
\label{subsec:Discord_arbitrary}

We consider the matrix of spins given in the Appendix by
Eq. (\ref{eq:DensityMatrix_spins_boosted}).
After projecting matrices as it is described above and 
finding their entropies and simplification of the resulting expressions, 
the combined entropy of  Eq. (\ref{eq:CombinationOfEntropies}) takes the 
form of a ratio of two fourth-order polynomials by $t=\tan(\theta/2)$:
\begin{equation}
\sum_s p_s E( \hat \varrho^{\Lambda}_{\sigma_A |\hat \Pi^{\sigma_B}_{s} })
= \frac{P_A(t, \varphi)}{P_B (t, \varphi)} = R(t, \varphi),
\label{eq:CombinedEntropy_arbitrary}
\end{equation}
where the coefficients of the polynomials of $P_A$ are functions
of $\varphi$, $\alpha$, $\beta$, $\delta$ and can be written in the form
\begin{eqnarray}
&&A_4 = A_0 = c_3^2 \sin^2 2\alpha  [c_1^2+c_2^2-(c_1^2-c_2^2)^2],
\label{eq:A4}
\\
&&A_3 = -A_1 = 4 c_3^3 \sin^2 2 \alpha \cos 2 \alpha \cos \varphi \times
\nonumber
\\
&& \qquad (c_1+c_2)^2(c_1-c_2),
\label{eq:A3}
\\
&&A_2 = 8 c_1 c_2 c_3^2 \sin^2 2\alpha \sin^2 \varphi + 
2 c_3^2 \sin^2 2 \alpha \times
\nonumber
\\
&& \qquad [(c_1^2 -c_2^2)^2 - 2c_1c_2],
\label{eq:A2}
\end{eqnarray}
and similarly, the coefficients of $P_B$,
\begin{eqnarray}
&& B_4 = B_0 = (c_1^2+c_3^2)(c_2^2 +c_3^2),
\label{eq:B4}
\\
&& B_3 = -B_1 = 2c_3 \cos 2 \alpha \cos \varphi \times 
\nonumber
\\
&& \qquad (c_1+c_2)^2(c_1-c_2),
\label{eq:B3}
\\
&& B_2 = c_1^4 + c_2^4 + 2c_3^4 + 2 c_3^2 \times
\nonumber
\\
&& \qquad [c_1^2 +c_2^2 - 2 \cos^2 2\alpha \cos^2
\varphi (c_1+c_2)^2].
\label{eq:B2}
\end{eqnarray}

\begin{center}
\begin{figure}
\includegraphics[width=\columnwidth]{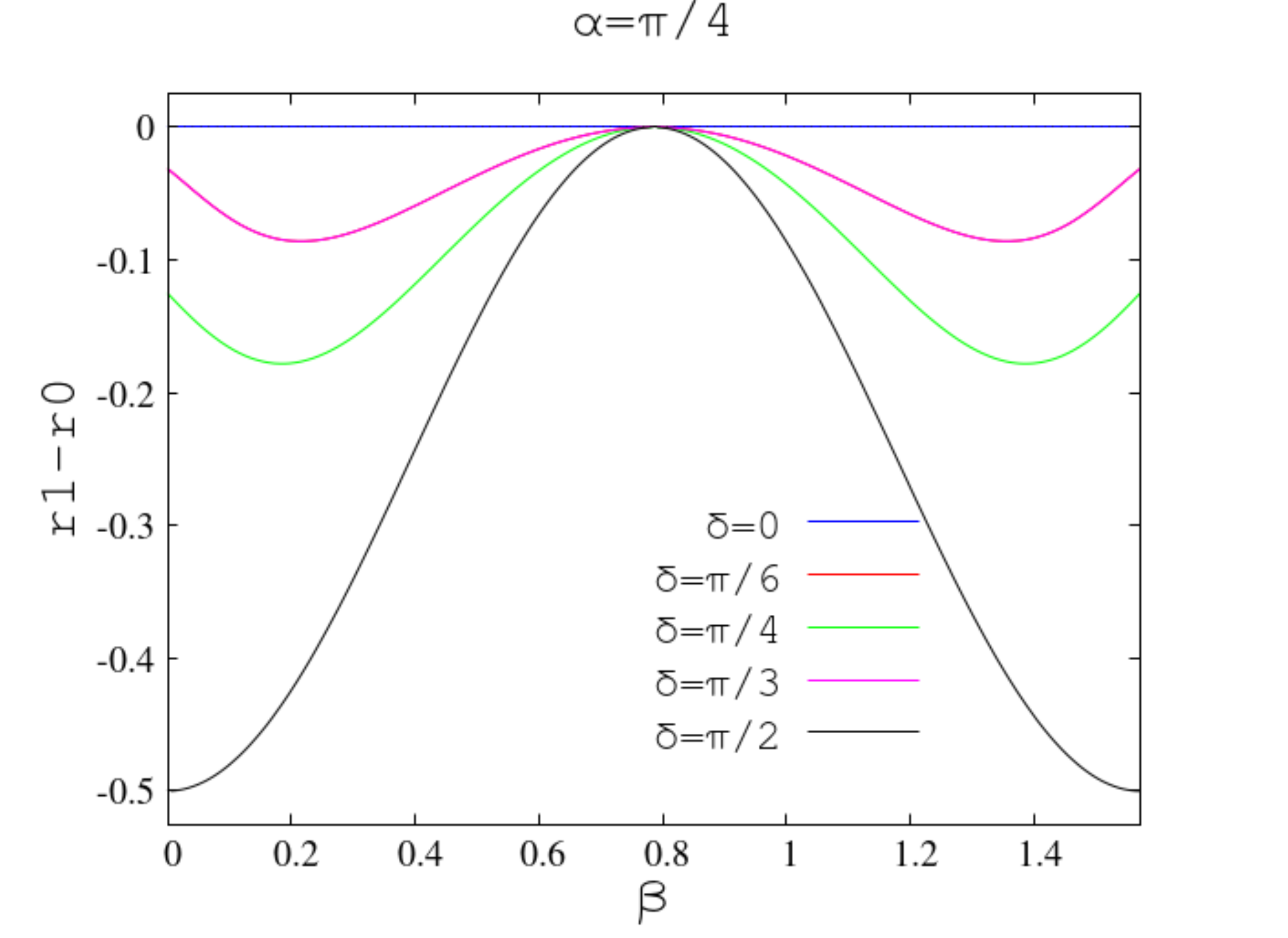}
\caption{Difference between the measurement entropies $r_1$ 
Eq. (\ref{eq:Rt1}) and $r_0$, Eq. (\ref{eq:Rt0}), corresponding
to $t=1$ and $t=0$, where the ratio of polynomials $R(t, \varphi)$
reaches its extremal values for $\alpha=\pi/ 4$.
It is seen $r_1 \leq r_0$ as $0 \leq \delta \pi/2$
and the difference reaches its maximum when 
$\delta=\pi/2$.
}
\label{fig:r10_pi4}
\end{figure}
\end{center}

\begin{center}
\begin{figure}
\includegraphics[width=\columnwidth]{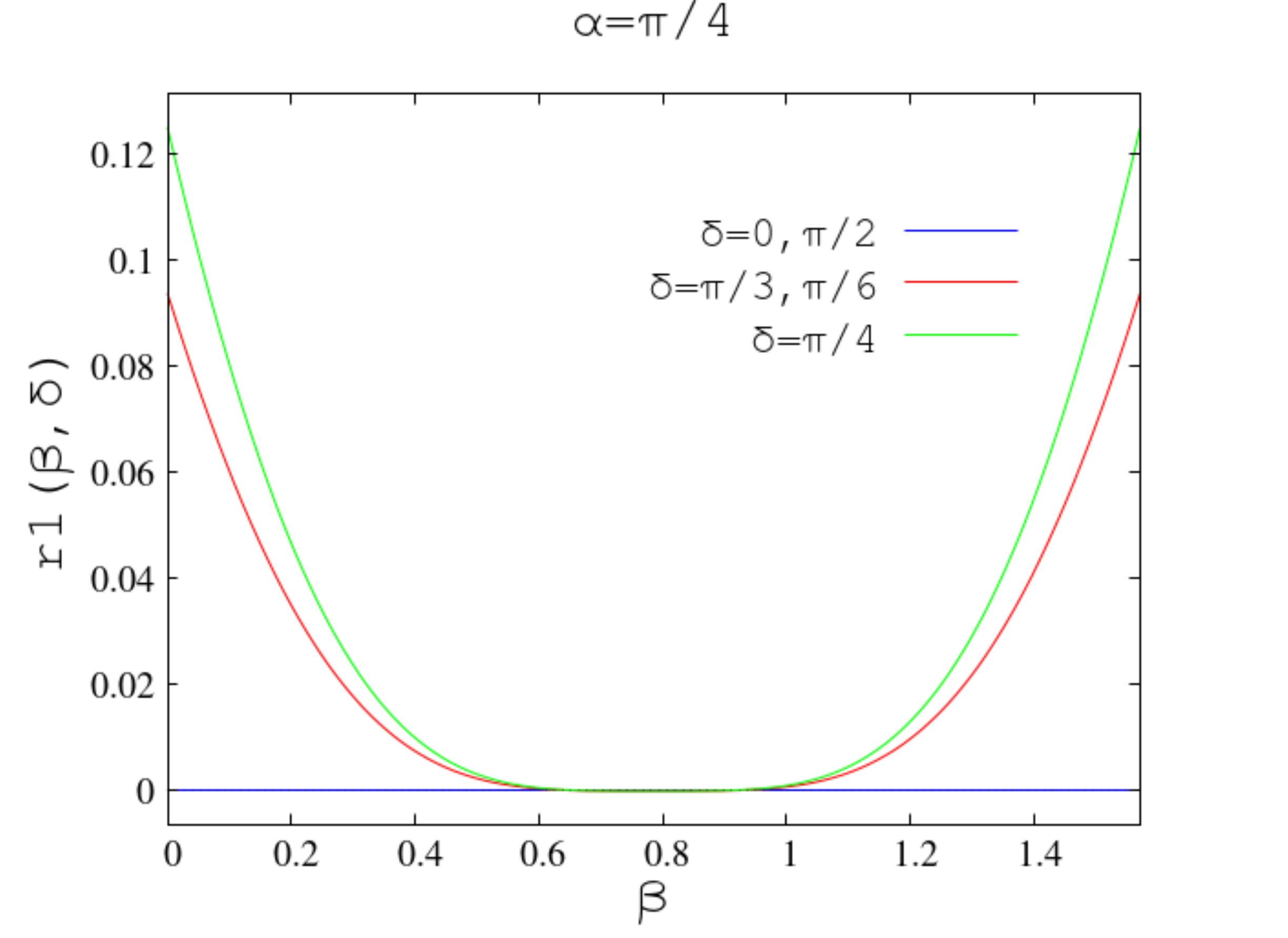}
\caption{$\beta$-dependence of the function $r_1$
for different values of $\delta$.}
\label{fig:r1bd_pi4}
\end{figure}
\end{center}

There exist a relatively simple analytically solvable
case when $\alpha=\pi/4$ which corresponds to maximal
mixing of momentum degrees of freedom.
Then it is easy to see that the minimal value of $R(t,\varphi)$ 
by $\varphi$ is obtained when $\sin \varphi=0$.
After taking derivative of $R(t,0)$ by $t$, the extremal values, where
the derivative becomes $0$, are $t=0$ and $t=1$, which correspond
to $\theta=0$ and $\theta=\frac \pi 2$, respectively.
The ratio of the polynomials depending on $\beta$ and $\delta$ can be 
expressed for $t=0, 1$ by respective functions:
\begin{eqnarray}
&& r_0(\beta, \delta)=\frac{A_4}{B_4} =  \sin^2 \delta (1-\sin 2 \beta) \times
\nonumber
\\
&& \qquad \qquad \left[ 1 - \frac 1 2  \frac{\sin^2 \delta(1-\sin 2\beta)}
{1-\cos^2 \delta \cos^2 2\beta}\right],
\label{eq:Rt0}
\\
&& r_1(\beta, \delta) = 
\frac{2A_4+A_2}{2B_4+B_2} = \frac {\sin^2\delta \cos^2 \delta} 2 \times
\nonumber
\\
&& \qquad \qquad (1-\sin 2\beta)^2.
\label{eq:Rt1}
\end{eqnarray}
It is seen that both functions are equal to zero before the boost $\delta=0$ 
and $r_1=0$ when $\delta =\pi/2$.
The equality of the minimum combined 
entropy to zero holds for $\delta=0, \pi/2$ for any $\alpha \neq \pi/4$ as 
it can be proven from Eqs. (\ref{eq:A4})-(\ref{eq:B2}) based on relations
with coefficients $c_i$, $i=1,2,3$, Eqs. (\ref{eq:crel1})-(\ref{eq:crel5}).
Fig. (\ref{fig:r10_pi4}) shows the dependence of the difference
$r_1-r_0$ on $\beta$ for different values of boost $\delta$ at fixed
$\alpha = \frac \pi 4$.
It becomes $0$ as $\delta \rightarrow 0^+$ and always negative for all
values of $\delta$ and $\beta$, therefore the choice $t^2=1$ 
is optimal for all $\delta \neq 0$ and $\beta$.
The dependence of $r_1$, which is the minimized combined entropy, 
on $\beta$ and $\delta$ is plotted in 
Fig. \ref{fig:r1bd_pi4}.
The figure demonstrates the symmetry to $\delta \rightarrow \pi/2-\delta$
which is evident from Eq. (\ref{eq:Rt1}).


Comparing Eq. (\ref{eq:Rt1}) with (\ref{eq:CombinationOfEntropies})
and (\ref{eq:QuantumDiscordGeneralExpression}), one finds that 
\begin{equation}
D_\Lambda\left(\alpha=\frac \pi 4, \beta, \delta \right)=\frac{\sin^2 2\beta}2,
\label{eq:Discord_pi4}
\end{equation}
when $\alpha=\pi/4$ for any $\beta$ and $\delta$.

The discord for other values of $\alpha$ including those different from $\pi/4$,
is found by numerical methods.
First note that the derivative by $\varphi$ from $R(t, \varphi)$ based on the
form of the coefficients of polynomials $P_A$ and $P_B$ it is possible
to prove that the minimum condition $\sin \varphi=0$ holds of all 
values of $\alpha$, $\beta$ and $\delta$.
This implies $\cos \varphi=\pm 1$, the plus sign corresponding to the minimum
and the minus to the maximum by $\varphi$.

\begin{center}
\begin{figure}
\includegraphics[width=\columnwidth]{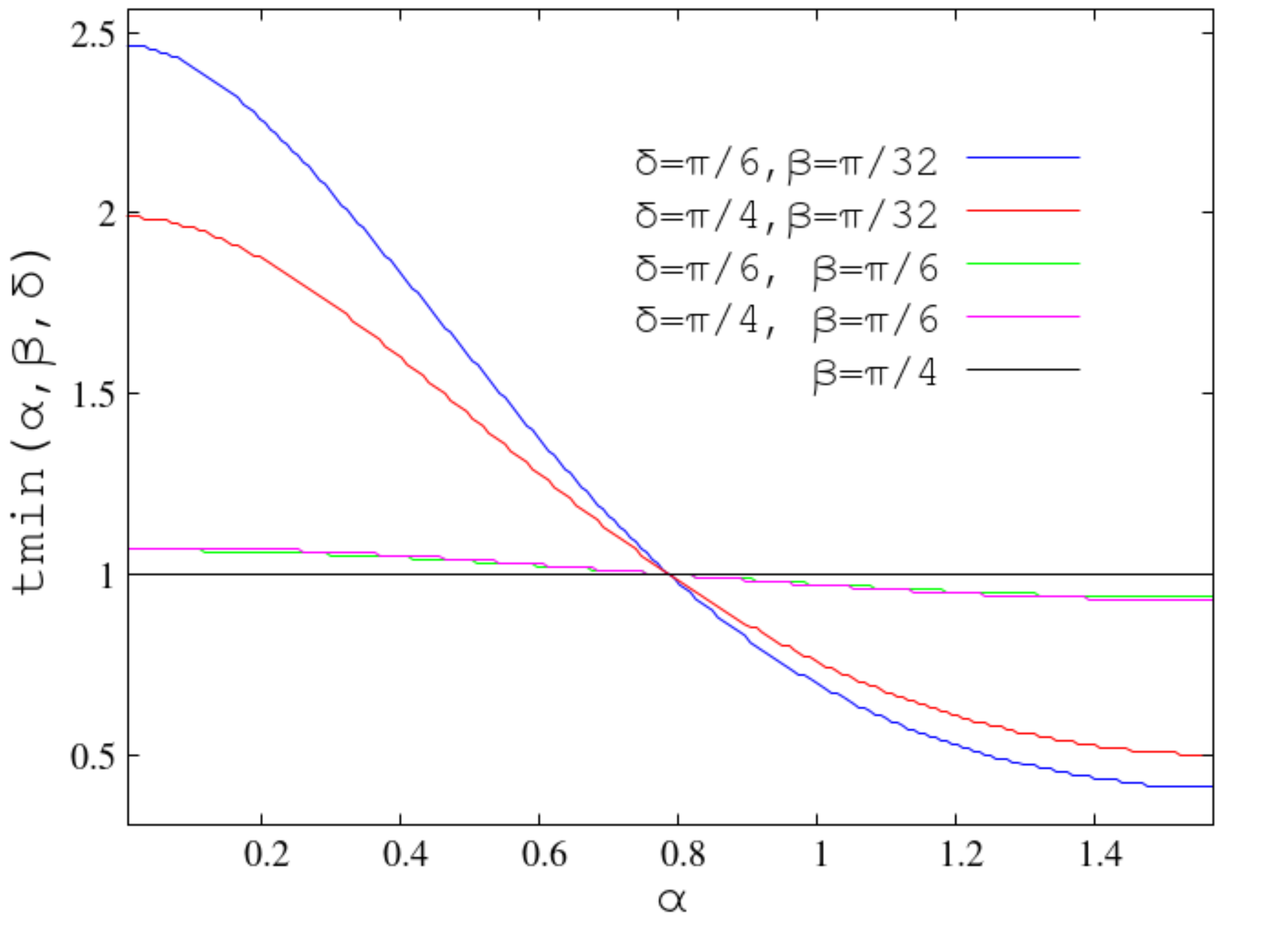}
\caption{$\alpha$-dependence of the $t_{min}$ minimizing the entropy
for different values of the parameters $\beta$ and $\delta$.
For every pair there exist a trajectory on the plot.
All trajectories cross $t=1$ as $\alpha=\pi/4$.}
\label{fig:roots_bd}
\end{figure}
\end{center}

Figure \ref{fig:roots_bd} depicts the dependence of the value
$t_{min}$ which minimizes the ratio of polynomials $P_A/P_B$ for different 
pairs of values of $\beta$ and $\delta$.
Every pair $(\beta, \delta)$ is characterized by a trajectory.
All trajectories pass through $t=1$ as $\alpha=\pi/4$ in
agreement with our previous discussion.
The trajectories are strongly $\delta$-dependent at small $\beta$ but the
dependence becomes weaker at intermediate values and it becomes a constant
$t=1$ when $\beta=\pi/4$.

\begin{center}
\begin{figure}
\includegraphics[width=\columnwidth]{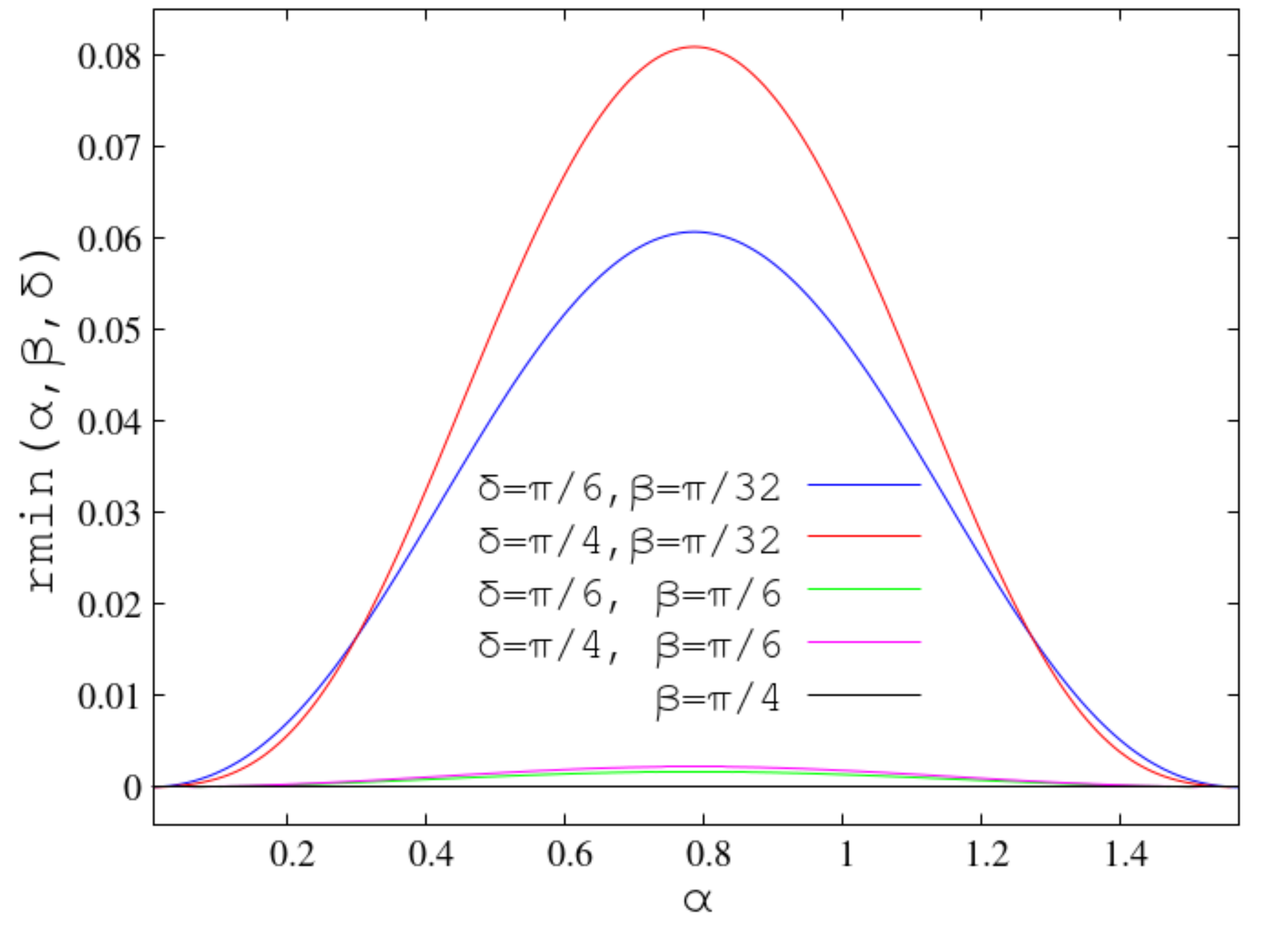}
\caption{$\alpha$-dependence of the minimum of combined entropy for
different values of $\beta$ and $\delta$.}
\label{fig:rmin_bd}
\end{figure}
\end{center}

Figure \ref{fig:rmin_bd} represents the dependence on $\alpha$ of the minima 
of the combined entropy for different values of $\beta$ and $\delta$.
The minimal entropy is identically zero at $\beta=\pi/4$ and rises at small
$\beta$ reaching maximum value at $\delta=\pi/4$.

\begin{center}
\begin{figure}
\includegraphics[width=\columnwidth]{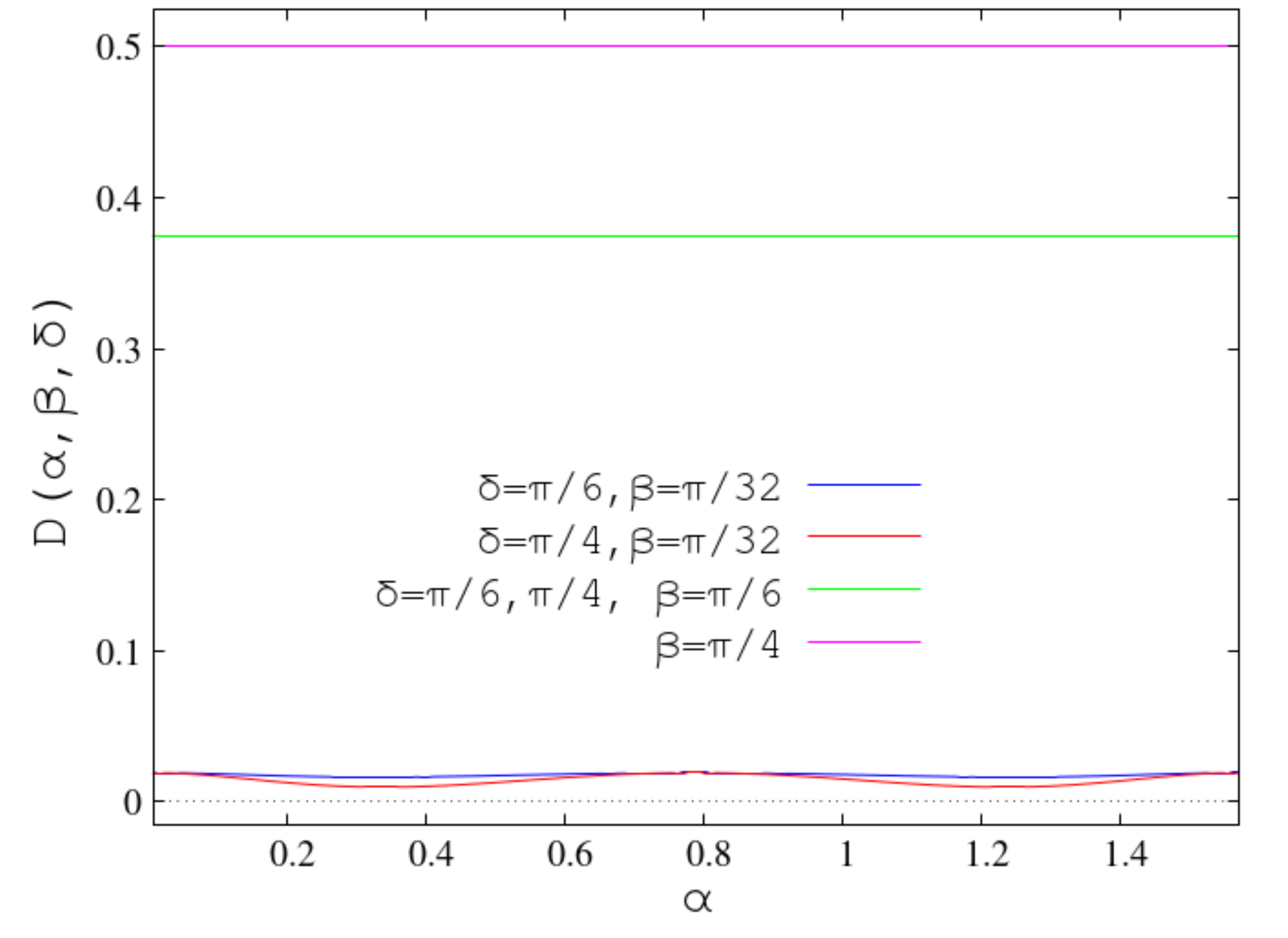}
\caption{$\alpha$-dependence of the discord for
different values of $\beta$ and $\delta$.}
\label{fig:dabd_a}
\end{figure}
\end{center}

\begin{center}
\begin{figure}
\includegraphics[width=\columnwidth]{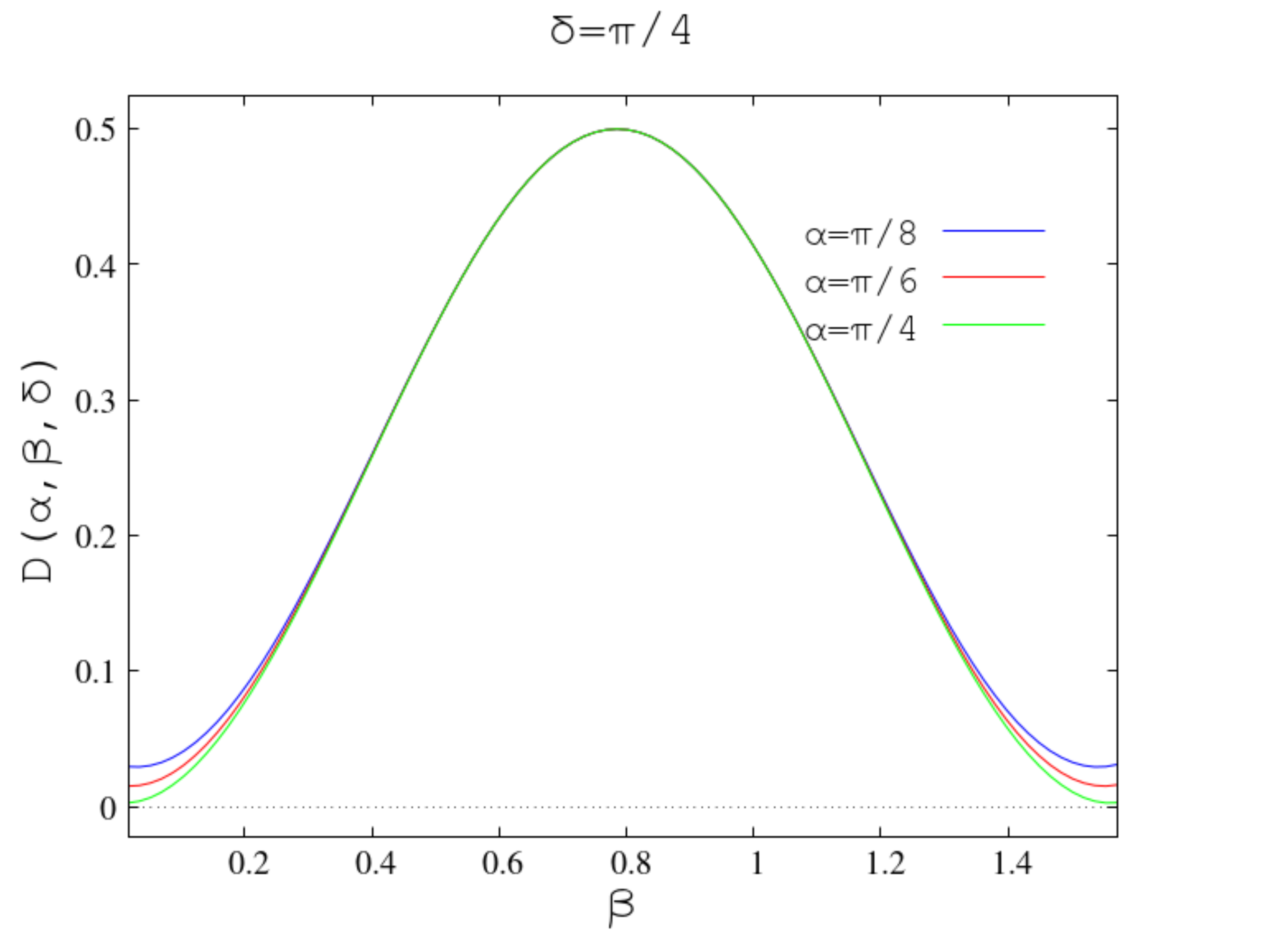}
\caption{$\beta$-dependence of the discord for different values
of $\alpha$ and fixed $\delta=\pi/4$.
The discord is approximately equal to $\sin^2(2\beta)/2$, analytical results
for $\alpha=\pi/4$, except of the ends of the interval
$\beta=0$ and $\beta=\pi/2$.}
\label{fig:dabd_b}
\end{figure}
\end{center}

The dependence of the quantum discord on $\alpha$ for several values of
$\beta$ is demonstrated in Fig. \ref{fig:dabd_a} and
the projection on $\beta$-axis is shown in Fig. \ref{fig:dabd_b}.
The maximum variation is observed at $\beta=0,\pi/2$, for
intermediate $\beta$-s it is approximately equal to 
$\sin^2(2\beta)/2$ for intermediate values of $\alpha$.

\subsection{Quantum discord after the measurement}
\label{subsec:discord_measured}

The concept of quantum correlations and entanglement is related to quantum 
coherence and measurements. 
Typically, measurements performed on a quantum 
system involve classical devices, and therefore the measurement setup 
exerts an unavoidable invasive effect on the coherence of the quantum state 
\cite{PhysRevLett.117.050402,PhysRevA.97.052101,emary2013leggett,
hoffmann2018structure,PhysRevA.87.052115,
PhysRevA.98.050302,PhysRevA.100.042103,PhysRevA.100.062314,
PhysRevLett.100.026804,PhysRevX.5.011003}. 
In order to explore the robustness of quantum discord with respect to
measurements for a particle-antiparticle pair, we evaluate quantum discord 
after implementing a POVM. 

Specifically, we consider the case when before and after the boost we measure
the spin polarization of one of the particles, chosen without loss of 
generality that corresponding to Alice.
We denote the wave function of the bipartite system before measurement as 
$|\phi\rangle$ for both the boosted and unboosted case. 
An efficient quantum measurement of spin polarization transforms this state i
into the post-measurement state
\begin{equation}\label{2postdensity matrix}
\hat\varrho_{\sigma_{Az}\sigma_B}=
\sum_{i=\pm}(\hat{\Pi}_{i}\bigotimes \hat{I}^{(B)})
\hat{\varrho}(\hat{\Pi}_{i}\bigotimes \hat{I}^{(B)}),
\end{equation}
where $\hat{\varrho}=|\phi \rangle\langle\phi \vert$, $\hat{I}^{(B)}$ is the
identity operator acting on the antiparticle $B$ spin space, and 
$\hat{\Pi}_{\pm}$ projects onto the $\pm$ spin component  along the $z$-axis.

\begin{eqnarray}
\label{post-measurement01}
\Phi \big\rangle=\frac{\big(\hat{\Pi}_{\pm}\bigotimes \hat{I}^{(B)}\big)
\big|\phi \big\rangle}{\sqrt{\big\langle 
\phi \big|\big(\hat{\Pi}_{\pm}\bigotimes
\hat{I}^{(B)}\big)\big|\phi \big\rangle}},
\end{eqnarray}
the positive/negative-helicity projector operator:
\begin{equation}
\hat{\Pi}_{\pm} = \frac{1 \pm \hat{p}\cdot \mathbf{\sigma}}{2}.
\label{eq:p_projection_spin_plus}
\end{equation}

For our case
The explicit expression for the post-measurement spin density matrix is 
given in Appendix \ref{sec:DerivationMeasurement}.
Based on it we compute the one-particle
\begin{equation}
E(\hat\varrho^\Lambda_{\sigma_{B}})= 
\frac 1 2(1-\cos^2\delta\cos^22\beta)
- \frac 1 2 \sin^2\delta \cos^2 2\beta \cos^2 2\alpha,
\label{eq:Entropy_Az_boosted_new}
\end{equation}
and two-particle spin entropies: 
\begin{widetext}
\begin{equation}
E(\hat \varrho^\Lambda_{\sigma_{Az}\sigma_B}) = 
\frac 1 2 (1-\cos^2\delta\cos^22\beta) +
\frac 1 2 \sin^2\delta \sin^22\alpha(1-\sin2\beta)
\left[1-\frac 1 2 \sin^2\delta(1-\sin2\beta)\right],
\end{equation}
and the discord reads
\begin{equation}
D_\Lambda (\sigma_{A_z}, \sigma_B) =
-\frac 1 2 \sin^2 \delta (1 -\sin 2\beta) 
\left[ 1+\sin 2\beta -\sin^2 2 \alpha
(\sin 2\beta + \frac {\sin^2\delta} 2 [1-\sin 2\beta])\right] +
\min\{\sum_j p_j E(\hat \varrho_{\sigma_{A_z}\vert \hat \Pi_j^B})\}.
\label{eq:Discord_measured_boosted}
\end{equation}
\end{widetext}

\begin{centering}
\begin{figure}
\includegraphics[width=\columnwidth]{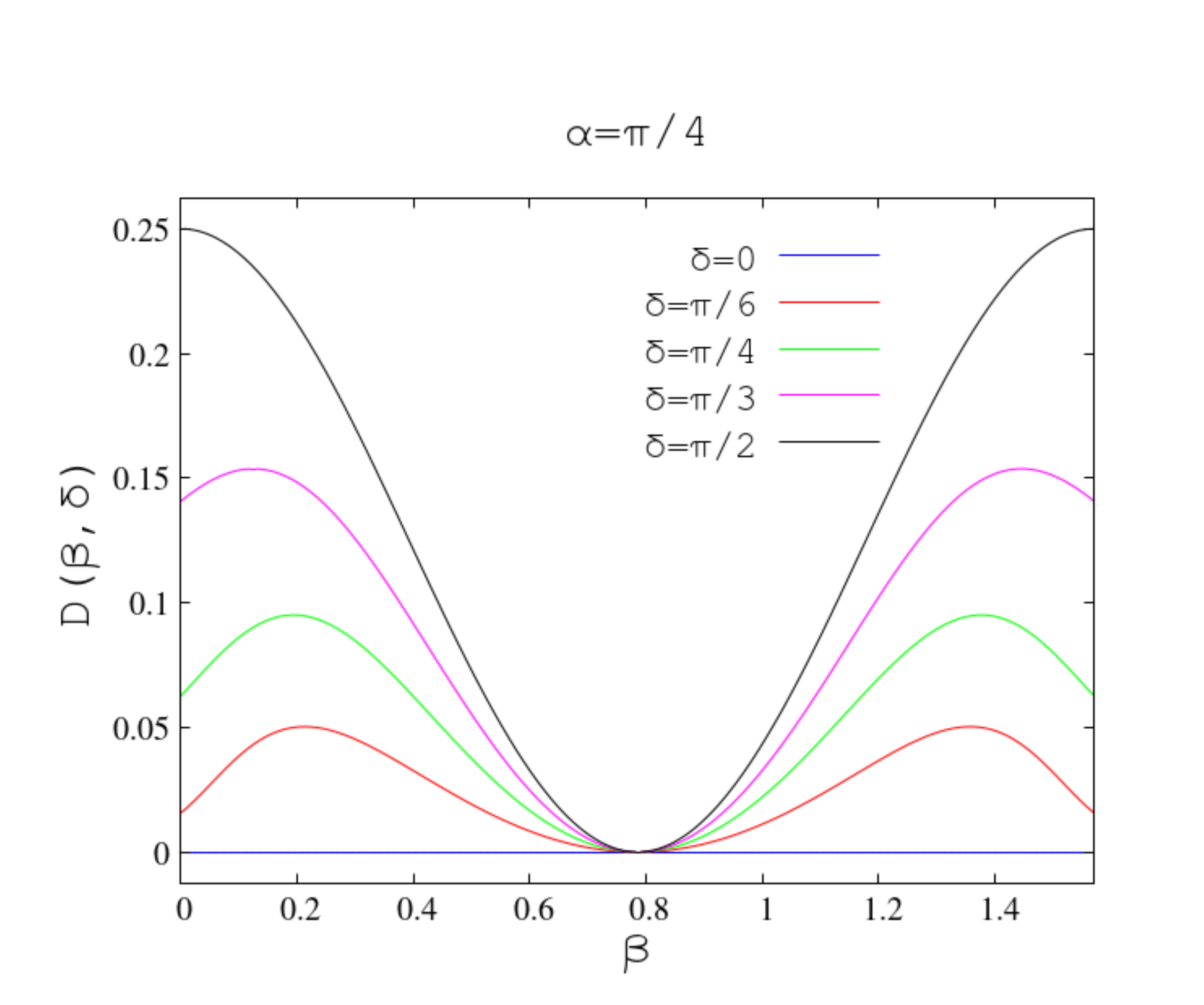}
\caption{Post-measurement quantum discord as a function of $\beta$ for
different choices of $\delta$ and fixed $\alpha=\pi/4$.}
\label{fig:dbd_pm}
\end{figure}
\end{centering}

\begin{centering}
\begin{figure}
\includegraphics[width=\columnwidth]{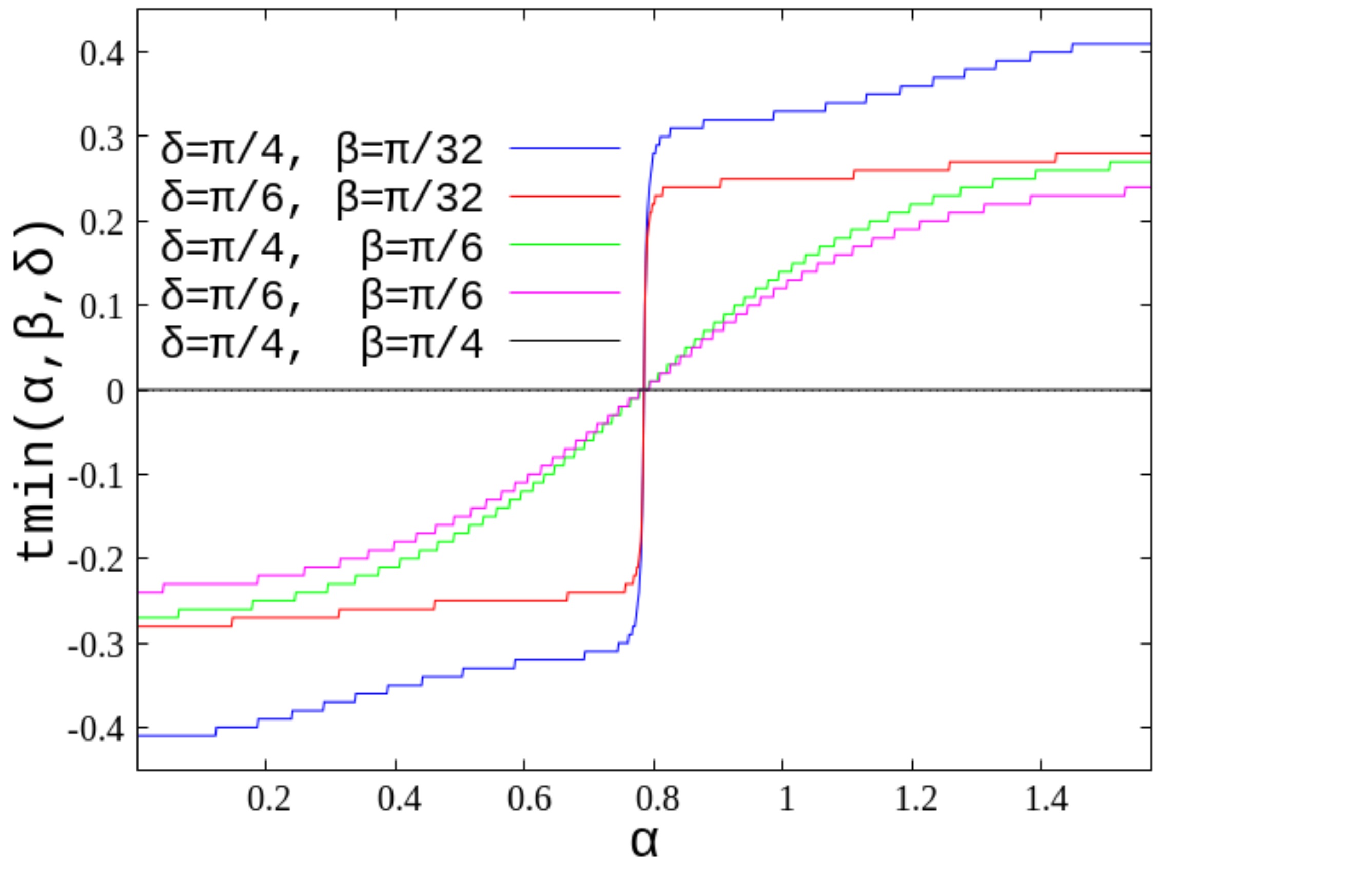}
\caption{$t_{min}$ minimizing the combined entropy
vs $\alpha$ for several pairs of values of the parameters $\beta$ and $\delta$.
The trajectories go through $t=0$ as $\alpha=\pi/4$.}
\label{fig:roots_bd_pm}
\end{figure}
\end{centering}

Calculation of the combined entropy follows the same lines as that for the case
before measurement.
The coefficients of the polynomials in the ratio similar to
Eq. (\ref{eq:CombinedEntropy_arbitrary}) are
\begin{eqnarray}
&& A_4=A_0=c_3^2[c_1^2+c_2^2-(c_1^2-c_2^2)^2],
\label{eq:A4_PM}
\\
&& A_3=-A_1 = 4 c_3\cos2\alpha \cos \varphi (c_2-c_1)\times
\nonumber
\\
&& \qquad [c_3^4 -(c_1+c_2)^2 c_3^2 -c_1^2c_2^2],
\label{eq:A3_PM}
\\
&& A_2 = 
2c_3^4[1-4 (c_1^2+c_2^2) \cos^2 2\alpha \cos^2\varphi  ] +
\nonumber
\\
&& \qquad 2c_3^2[ (c_1^2-c_2^2)^2 - 8c_1^2c_2^2\cos^2 2\alpha \cos^2\varphi] +
\nonumber
\\
&& \qquad \qquad 2c_1^2c_2^2,
\label{eq:A2_PM}
\end{eqnarray}
and
\begin{eqnarray}
&& B_4=B_0 = (c_1^2+c_3^2)(c_2^2+c_3^2),
\\
&& B_3=-B_1 = 2c_3 \cos2\alpha\cos\varphi \times
\nonumber
\\
&& \qquad \qquad (c_1+c_2)^2 (c_1-c_2),
\label{eq:B3_PM}
\\
&& B_2= c_1^4+c_2^4 + 2c_3^4 + 2c_3^2 \times
\nonumber
\\
&&\qquad [c_1^2+c_2^2 -
2\cos^2 2\alpha \cos^2 \varphi (c_2+c_1)^2].
\label{eq:B2_PM}
\end{eqnarray}
As before, one can observe that the minimum value by $\varphi$ corresponds
to $\sin \varphi=0$, $\cos \varphi=1$.
A simplest case where the model can be solved analytically, again corresponds
to $\alpha=\pi/4$.
Then analogously to Eq.(\ref{eq:Rt0}) and (\ref{eq:Rt1}), the extremal values
by $t$ are at $t=0,1$ with corresponding functions
\begin{eqnarray}
&& r_0(\beta, \delta)= \sin^2 \delta (1-\sin 2 \beta) \times
\nonumber
\\
&& \qquad \qquad \left[ 1 - \frac 1 2  \frac{\sin^2 \delta(1-\sin 2\beta)}
{1-\cos^2 \delta \cos^2 2\beta}\right],
\label{eq:Rt0_pm}
\\
&& r_1(\beta, \delta) = \frac 1 2 (1-\cos^2\delta\cos^22\beta).
\label{eq:Rt1_pm}
\end{eqnarray}
Unlike the previous case, now $r_1>r_0$ for all values of
$\beta$ and $\delta$.
Another important difference from the pre-measurement case is that the minimal
value of the combined entropy is equal to $0$ only for $\delta=0$, and it rises
as $\delta$ goes to $\pi/2$.
The corresponding discord has the form:
\begin{widetext}
\begin{equation}
D_\Lambda\left(\alpha=\frac \pi 4,\beta,\delta \right) 
=\frac 1 2 \sin^2\delta (1-\sin2\beta)
\left[1- \frac 1 2 \sin^2\delta (1-\sin2\beta)
\frac{1+\cos^2\delta\cos^22\beta}{1-\cos^2\delta\cos^22\beta}\right].
\label{eq:Discord_pi4_pm}
\end{equation}
\end{widetext}
The dependence of the discord on $\beta$ and $\delta$ is shown in
Fig. (\ref{fig:dbd_pm})
The values of $t$ minimizing the combined entropy for any $\alpha\neq\pi/4$ and 
different pairs of $\beta$ and $\delta$ are shown in 
Fig. \ref{fig:roots_bd_pm}.
The trajectories cross each other at $t=0$ when $\alpha=\pi/4$.

\begin{centering}
\begin{figure}
\includegraphics[width=\columnwidth]{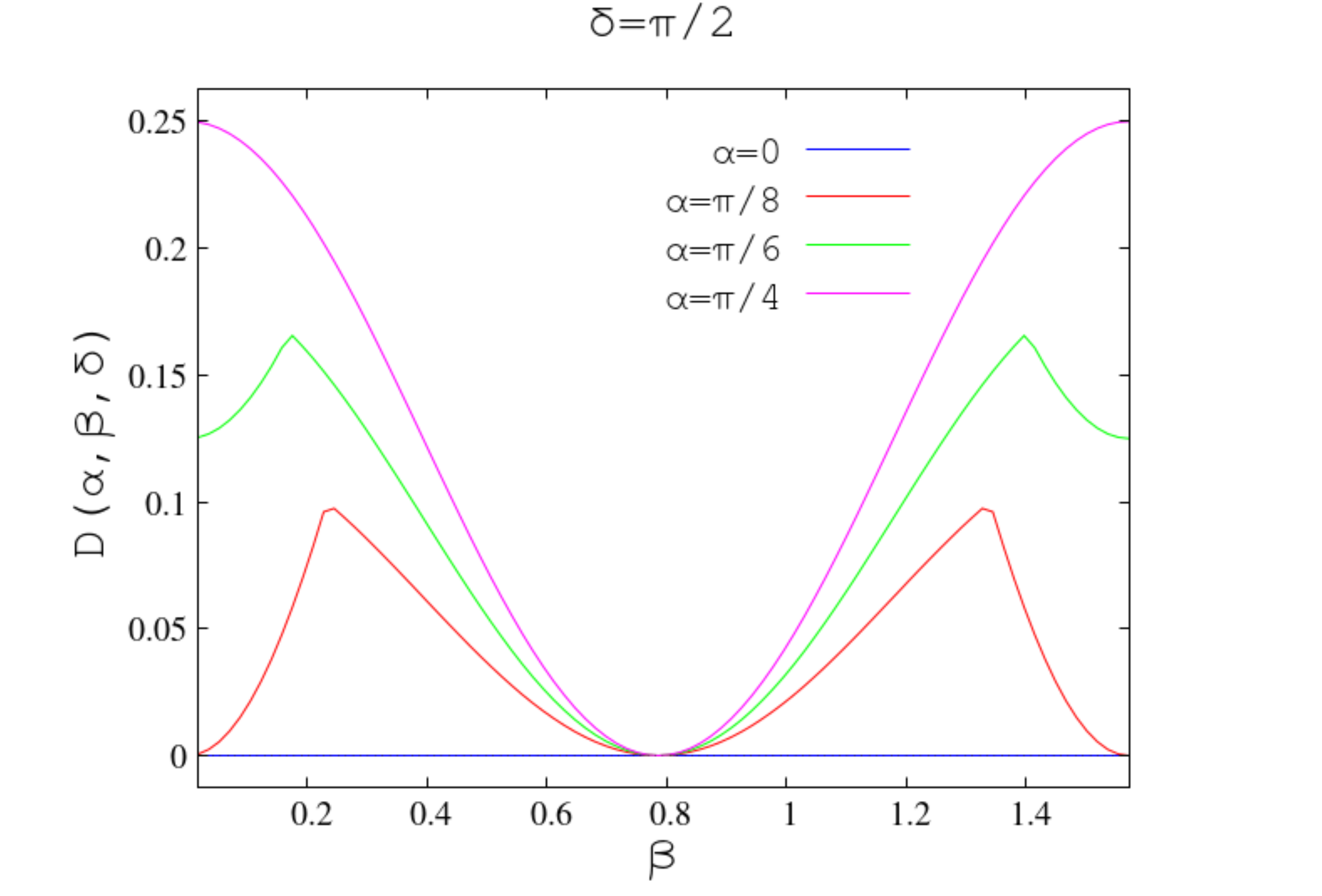}
\label{fig:dab_pi2_pm}
\caption{Dependence of the discord for post-measurement spin density matrix
on $\beta$ for different $\alpha$-s at $\delta=\pi/2$.}
\end{figure}
\end{centering}

The discord dependence on $\beta$ for different $\alpha$-s
is presented in Fig. \ref{fig:dab_pi2_pm}

\section{Experimental remarks}\label{sec:Experiment}

\subsection{Top-antitop quark production}

We finally discuss real high-energy processes, relating them to the results of this work. We focus on the particular case of a top/antitop ($t\bar{t}$) pair, quite unique in high-energy physics because of its large mass (indeed, the top quark is the most massive fundamental particle in the Standard Model). This large mass is translated into a large decay width that makes each one of top/antitop in the $t\bar{t}$ pair to decay well before any other process, such as hadronisation or spin decorrelation, can affect the $t\bar{t}$ spins. As a consequence, the information regarding the spins of the $t\bar{t}$ pair is inherited uncorrupted by the decay products. 

Specifically, the decay spin density matrix of the top quark, describing the decay of the top quark to some final state $F$, is defined as
\begin{equation}
    \Gamma_{\sigma'\sigma}\equiv \bra{F}T\ket{t\sigma}\bra{t\sigma'}T^{\dagger}\ket{F}
\end{equation}
where $\ket{t\sigma}$ is a top quantum state with spin $\sigma$, and $T$ the \textit{on-shell} $T$-matrix of the decay process. A similar decay spin density matrix $\bar{\Gamma}_{\sigma'\sigma}$ can be defined for the decay of an antitop quark to a state $\bar{F}$. We restrict to the case of a dileptonic decay of the $t\bar{t}$ pair
\begin{eqnarray}\label{eq:ttdecays}
    t&\rightarrow& b+l^+ + \nu_l,\\
\nonumber \bar{t}&\rightarrow& \bar{b}+l^{-}+\bar{\nu}_l
\end{eqnarray}
If we switch to the top/antitop rest frames and integrate all the degrees of freedom of the final states except for the antilepton/lepton directions, due to rotational invariance, the spin decay density matrices take the simple form \cite{Baumgart2013}
\begin{equation}\label{eq:DecaySpinDensityMatrix}
    \Gamma \propto \frac{\sigma^0+\hat{\mathbf{\ell}}_{+}\cdot \mathbf{\sigma}}{2},~
    \bar{\Gamma}=\frac{\sigma^0-\hat{\mathbf{\ell}}_{-}\cdot \mathbf{\sigma}}{2}
\end{equation}
$\hat{\mathbf{\ell}}_{\pm}$ being the antilepton/lepton directions. 

With the help of the decay density matrices, the angular differential cross-section characterizing the dileptonic decay of a $t\bar{t}$ pair is computed in the so-called narrow-width approximation \cite{Bernreuther1994} as
\begin{equation}\label{eq:LeptonicDifferentialCrossSectionTotal}
 \frac{\mathrm{d}\sigma_{\ell \bar{
 \ell}}}{\mathrm{d}\Omega_{+}\mathrm{d}\Omega_{-}} \sim \int\mathrm{d}M\mathrm{d}\Omega~\textrm{Tr}\left[\Gamma R(M,\hat{p}) \bar{\Gamma}\right]
\end{equation}
where we are integrating over all possible values of the $t\bar{t}$ energy and momentum $M,\hat{p}$. Since after integrating over all lepton/antilepton directions, $\Gamma,\bar{\Gamma}\sim \sigma^0$, the total dileptonic cross section $\sigma_{\ell\bar{\ell}}$ is proportional to the integral of $\tilde{C}_{00}$. As a result, the normalized angular distribution reads
\begin{equation}\label{eq:LeptonicCrossSectionExperimentNormalized}
\frac{1}{\sigma_{\ell\bar{\ell}}}\frac{\mathrm{d}\sigma}{\mathrm{d}\Omega_{+}\mathrm{d}\Omega_{-}}=\frac{1+\mathbf{B}^+\cdot\hat{\mathbf{\ell}}_{+}-\mathbf{B}^-\cdot\hat{\mathbf{\ell}}_{-}-\hat{\mathbf{\ell}}_{+}\cdot \mathbf{C} \cdot\hat{\mathbf{\ell}}_{-}}{(4\pi)^2}
\end{equation}
The vectors $\mathbf{B}^{\pm}$ are the integrated top (antitop) spin polarizations and $\mathbf{C}$ is the integrated spin correlation matrix [see Eq. (\ref{eq:SpinCorrelations}) and ensuing discussion], where the integrated expectation value of an observable $O$ is
\begin{equation}
    \braket{O}=\frac{\int\mathrm{d}M\mathrm{d}\Omega~\textrm{Tr}\left[R(M,\hat{p})O\right]}{\int\mathrm{d}M\mathrm{d}\Omega~\textrm{Tr} R(M,\hat{p})}
\end{equation}
Interestingly, one can cut the integrals in both $M$ and $\hat{p}$ in a high-energy collider by reconstructing the $t\bar{t}$ momenta, restricting these expectations values to certain regions of phase space  \cite{Afik:2020onf,Afik:2022kwm}. In this way, one can in principle reconstruct the spin quantum state of the $t\bar{t}$ pair as given by its spin density matrix $\hat\varrho(M,\hat{p})$.

\subsection{Specific measurements with top-antitop quark pairs}

We now relate the results of this work with relevant experimental observables for $t\bar{t}$ pairs at the LHC. For that purpose, we compare the wave function of Eq. (\ref{eq:Wavefunction_separable_form}) with the spin density matrix $\hat\varrho(M,\hat{p})$, accessible in experiments.

Regarding the momentum part of the wave function, we note that $\hat\varrho(M,\hat{p})$ already describes the result of a POVM of the momentum of any particle of the pair. Indeed, the momenta of the $t\bar{t}$ pair can be measured on an event by event basis. Moreover, regardless of the direction $\hat{p}$, due to the properties of $t\bar{t}$ QCD production, at the LHC it can only be $\alpha=\pm \pi/4$ \cite{Afik:2020onf,Afik:2022kwm}.

Regarding the spin part of the wave function of Eq. (\ref{eq:Wavefunction_separable_form}), in general $\hat\varrho(M,\hat{p})$ is not a pure state. However, for $t\bar{t}$ production at the LHC,  close to threshold or at high-$p_T$, $\hat\varrho(M,\hat{p})$ is to a very good approximation a singlet ($\beta=-\pi/4$) or triplet ($\beta=\pi/4$) pure state, respectively \cite{Afik:2020onf,Afik:2022kwm}. In contrast to the case of the momentum, directly measurable, the $t\bar{t}$ spin correlations and polarizations are only obtained \textit{a posteriori} by fitting the cross-section of Eq. (\ref{eq:LeptonicCrossSectionExperimentNormalized}). We stress that the appealing distribution of the decay spin density matrix of Eq. (\ref{eq:DecaySpinDensityMatrix}) only arises after integrating over all the remaining degrees of freedom. Thus, spins cannot be directly measured and the POVM formalism cannot be applied to them.

As a result of the above considerations, we conclude that the result of any POVM that does not involve at least the momentum of one of the quarks cannot be implemented in a high-energy collider. For the partitions used throughout this work, this implies:

\begin{itemize}
\item All density matrices of the $1+3$ partitioning can be obtained from the production spin density matrix.

\item Regarding the $p+\sigma$ partitioning, only the density matrix after tracing over momentum can be measured.

\item The density matrices of the $A+B$ partitioning can be also measured from the production spin density matrix.
\end{itemize}

Finally, we discuss the role of a Lorentz transformation. So far, we have analyzed the spin density matrix $\hat\varrho(M,\hat{p})$, which describes the $t\bar{t}$ momenta in the c.m. frame, as in the wave function of Eq. (\ref{eq:Wavefunction_momentum}), while the spins are described in their respective top (antitop) rest frames [see Eq. (\ref{eq:DecaySpinDensityMatrix}) and ensuing discussion], where spin is well defined.

In principle, since $t\bar{t}$ momenta are reconstructed from the directions of the decay products, they can be determined in any reference frame. However, within the current experimental scheme, the spin observables are always measured in the parent $t\bar{t}$ rest frames. Nevertheless, one could think about measuring the lepton directions in different frames. Future works should examine in detail how the dileptonic cross section of Eq. (\ref{eq:LeptonicDifferentialCrossSectionTotal}) behaves under a Lorentz transformation of the lepton/antilepton momenta, and which Lorentz transformed $t\bar{t}$ spin observables can be extracted.

\section{Conclusions}
\label{sec:Conclusions}

Relativistic effects may exert a specific impact on the quantumness of a system. The inherently quantum correlations are quantified through entropic entanglement measures and quantum discord. Inspired by this idea, we have studied the effect of a Wigner rotation in a particle-antiparticle pair. 

The relativistic principle is universal, and it requires Lorentz invariance. Thus, the quantum correlations stored in the entire system or each of its subsystems should be also invariant under Lorentz transformations. Nevertheless, while entropies of subsystems are indeed invariant $S\left(\hat \varrho_{A}\right)=S\left(\hat \varrho_{A}^\Lambda\right)=S\left(\hat \varrho_{B}\right)=S\left(\hat \varrho_{B}^\Lambda\right)$, one can find different types of partitions of the Hilbert space whose entropies are not Lorentz invariant. Here, we consider a bipartite system $\hat \varrho_{AB}$ whose subsystems are in turn composed by two subsystems, i.e., spin and momentum sectors: $\hat \varrho_{AB}=\hat \varrho_{\sigma_Ap_A\sigma_Bp_B}$. We trace the entire system over all possible partitions and apply relativistic boosts, computing
their entropy and the mutual information between the different parts of the system. 

Regarding quantum discord, we have shown that, depending on the initial state and the parameters of the boost, the discord of the boosted state can become quite large. We have calculated the difference between discords corresponding to the states before and after boost. We observe an interesting fact: an initially disentangled state with zero discord can become entangled after the boost. Another interesting fact concerns symmetries. 
We have also observed that quantum discord generated by Lorentz boost is 
robust concerning the protective POVM, while the same measurement exerts an 
invasive effect on the discord of the initial state. 

Finally, we have discussed how the results of this work could be 
measured using top quarks, opening the perspective to implement our 
scheme in a high-energy collider such as the LHC.

\acknowledgements

We are extremely grateful to Yoav Afik from Experimental Physics Department at
CERN 
and Juan Ramón Muñoz de Nova 
(Departamento de F\'isica de Materiales, Universidad Complutense de Madrid),
for very useful discussion of the theoretical part of the work and its
practical application.



\appendix
\numberwithin{equation}{section}
\section{Calculation of entropy for any density matrix}
\label{sec:CalculationOfEntropy}

We consider a multipartite quantum system consisting of $n$ subsystems.
Sets of orthonormal basis states $\psi^1_{i_1},\dots,\psi^n_{i_n}$
span the Hilbert spaces of $1,\dots,n$-th subsystem, respectively,
with indices $i_n$ running in the range
from $1$ to the dimension of the $n$-th subsystem.
Any state of the full quantum system can be expressed as follows
\begin{equation}
\ket{\Psi} = \sum_{i_1,\dots,i_n} c_{i_1,\dots,i_n} 
\ket{\psi^1_{i_1}}\otimes\dots\otimes\ket{\psi^n_{i_n}}.
\label{eq:Full_QuantumState}
\end{equation}
The density matrix can be written in the form
\begin{equation}
\hat \varrho = \ket{\Psi}\bra{\Psi} = 
\sum_{[i],[j]} \rho_{[i][j]} \ket{\psi_{[i]}}\bra{\psi_{[j]}}, 
\label{eq:Full_DensityMatrix}
\end{equation}
where $[i]$, $[j]$ denote sets of indices $_1,\dots,i_n$, $j_1,\dots,j_n$,
$\rho_{[i][j]}=c_{[i]}c^*_{[j]}$, and $\psi_{[i]}$, $\psi_{[j]}$ 
stand for direct products of basis states.
The reduced density matrix of the $k$-th subsystem is defined after taking the trace over the all other subsystems except of $k$-th:
\begin{equation}
\hat \varrho_k = \sum_{[i]/k} \bra{\psi_{[i]/k}}\hat \varrho \ket{\psi_{[i]/k}}
= \sum_{i_k j_k} \rho^k_{i_kj_k} \ket{\psi^k_{i_k}}\bra{\psi^k_{j_k}}.
\label{eq:DensityMatrix_reduced}
\end{equation}
Here $[i]/k$ in the above equations denotes the set of indices of all subsystems $i$ except those of the $k$-th subsystem, $i_k$.

The elements of the reduced density matrix 
$\hat \varrho_k$ are expressed in terms of the matrix elements of the full density matrix
Eq. (\ref{eq:Full_DensityMatrix}) 
by taking tracing all pairs of indices except  $\{i_k,j_k\}$:
\begin{equation}
\rho^k_{i_k j_k} =
\sum_{[i]/k} \rho_{[i]/k, [i]/k}.
\label{eq:DensityMatrix_reduced2}
\end{equation}
Now we calculate the trace of the reduced matrix $\varrho_k$ squared:
\begin{eqnarray}
&&\mathrm{Tr}\hat\varrho^2_k 
= \sum_i \bra{\psi^k_i}\hat \varrho_k^2\ket{\psi^k_i} =
\nonumber
\\
&& \sum_{i,j} \bra{\psi^k_i}\hat \varrho_k\ket{\psi^k_j} 
\bra{\psi^k_j}\hat \varrho_k\ket{\psi^k_i} = 
\sum_{i,j} \vert \rho^k_{ij}\vert^2,
\label{eq:Trace_ReducedMatrix}
\end{eqnarray}
since the density matrix is Hermitian and 
$\bra{\psi_i}\ket{\psi_j}=\delta_{ij}$ because of the orthonormality condition
for the basis functions.

After partitioning the full system into a set of subsystems, each subsystem is characterized by its reduced density matrix $\varrho^i$. Using Eq. (\ref{eq:Trace_ReducedMatrix}), it is shown that Eq. (\ref{eq:Entanglement_definition}) can be simply rewritten as Eq. (\ref{eq:Entanglement_definition_expanded}).


\section{Simplest case: a system of two spins}
\label{sec:SimplestCase}
For the sake of simplicity, we consider a bipartite system of two spins $A$ and $B$ shared between Alice and Bob.
The spin states are formed by the direct product of Alice's and Bob's single spins functions, and the complete basis for the two spin states is formed by the eigenstates of the $z$-spin component:
\begin{equation}
\{\ket{\psi_1}, \ket{\psi_2},\ket{\psi_3}, \ket{\psi_4}\} =
\{\ket{\uparrow \downarrow}, \ket{\downarrow \uparrow}, 
\ket{\uparrow \uparrow}, \ket{\downarrow \downarrow} \}.
\label{eq:Two_spin_basis}
\end{equation}
Any density matrix in this Hilbert space has the general form
\begin{equation}
\hat \varrho = \sum_{i,k=1}^4 \rho_{ik}\ket{\psi_i}\bra{\psi_k}.
\label{eq:DensityMatrix_Alice_Bob}
\end{equation}

We explicitly express the wave functions corresponding to definite spin orientation for both subsystems A and B
by tracing the basis vectors given in Eq. (\ref{eq:Two_spin_basis}):
\begin{align}
&\ket{\uparrow}_A =
\ket{\psi_1}  \bra{\downarrow}_B + \ket{\psi_3} \bra{\uparrow}_B,
\label{eq:Aup}
\\
&\ket{\downarrow}_A = 
\ket{\psi_2}  \bra{\uparrow}_B + \ket{\psi_4} \bra{\downarrow}_B,
\label{eq:Adown}
\\
&\ket{\uparrow}_B = 
\ket{\psi_2} \bra{\downarrow}_A
+ \ket{\psi_3} \bra{\uparrow}_A,
\label{eq:Bup}
\\
&\ket{\downarrow}_B =
\ket{\psi_1} \bra{\uparrow}_A + \ket{\psi_4} \bra{\downarrow}_A
.
\label{eq:Bdown}
\end{align}
The matrix of the subsystem of Alice is obtained by taking the trace over
the states of Bob using Eqs. (\ref{eq:Bup}) and (\ref{eq:Bdown}):
\begin{equation}
\hat \varrho_A = \mathrm{Tr}_B \hat \varrho = 
\begin{pmatrix}
\rho_{11}+\rho_{33} & \rho_{14}+\rho_{32}\\
\rho_{41}+\rho_{23} & \rho_{22}+\rho_{44}
\end{pmatrix}.
\label{eq:Subsystem_of_Alice}
\end{equation}
Bob's spin density matrix is obtained by exchanging $2 \leftrightarrow 1$. 

As a next step we measure the spin of particle $A$.
Eqs. (\ref{eq:Aup})-(\ref{eq:Bdown}) can be expanded into the following
form:
\begin{align}
\ket{\uparrow}\bra{\uparrow}_A & = \ket{\psi_1}\bra{\psi_1}
+ \ket{\psi_3}\bra{\psi_3},
\label{eq:Aup2}
\\
\ket{\downarrow}\bra{\downarrow}_A & = \ket{\psi_2}\bra{\psi_2} +
\ket{\psi_4}\bra{\psi_4},
\label{eq:Adown2}
\\
\ket{\uparrow}\bra{\uparrow}_B & = \ket{\psi_2}\bra{\psi_2} +
\ket{\psi_3}\bra{\psi_3},
\label{eq:Bup2}
\\
\ket{\downarrow}\bra{\downarrow}_B & = \ket{\psi_1}\bra{\psi_1} +
\ket{\psi_4}\bra{\psi_4}.
\label{eq:Bdown2}
\end{align}
The result of the measurement of the spin of particle $A$ in the basis of Eq. (\ref{eq:Two_spin_basis}) is given by 
the following matrix:
\begin{equation}
\hat\varrho_{\sigma_{Az}\sigma_B} =
\begin{pmatrix}
\rho_{11} & 0 & \rho_{13} & 0\\
0 & \rho_{22} & 0 & \rho_{24} \\
\rho_{31} & 0 & \rho_{33} & 0\\
0 & \rho_{42} & 0 & \rho_{44}
\end{pmatrix}.
\label{eq:DensityMatrix_AfterMeasurement_4x4}
\end{equation}
As one can see, only the diagonal elements and those containing pairs of
indices $(1,3)$ and $(2,4)$ are non-zero.
After performing the same measurement for the spin of particle $B$, one finds that the
non-zero elements correspond to the diagonal elements and to those with pairs
of indices $(1,4)$ and $(2,3)$. As a result of these measurements, an initially non-entangled state may become entangled after the measurement because it removes some matrix elements from the squared sum of Eq. (\ref{eq:Entanglement_definition_expanded}), which as a result it is no longer equal to $1$.

In this work we consider a mixed state of two antiparallel spins depending
on the angular parameter $\beta$:
\begin{equation}
\ket{\psi} = \cos \beta \ket{\uparrow \downarrow} +\sin \beta
\ket{\downarrow \uparrow}.
\label{eq:Wavefunction_beta_dependent}
\end{equation}
The density matrix
corresponding to the function in Eq. (\ref{eq:Wavefunction_beta_dependent})
has the form:
\begin{equation}
\hat \varrho_{\sigma_A\sigma_B} =
\begin{pmatrix}
c^2 && s\cdot c && 0 && 0\\
s \cdot c && s^2 && 0 && 0\\
0 && 0 && 0 && 0\\
0 && 0 && 0 &&0
\end{pmatrix},
\label{eq:Matrix_rho_alpha_default_basis}
\end{equation}
where $c=\cos \beta$, $s=\sin \beta$.
It is easy to see, that the sum of the squares of matrix elements is
equal to $1$, therefore the entanglement of the entire matrix is
zero.

Taking trace over the states of $B$,
\begin{equation}
\hat \varrho_{\sigma_A} =
\cos^2 \beta \ket{\uparrow}\bra{\uparrow}_A + 
\sin^2 \beta \ket{\downarrow}\bra{\downarrow}_A,
\label{eq:Subsystem_of_Alice2}
\end{equation}
and the associated contribution to the entanglement is
\begin{equation}
E(\hat\varrho_{\sigma_A}) = 1 - (\cos^4 \beta + \sin^4 \beta) = 
\frac {\sin^2 2 \beta} 2.
\label{eq:Alice_contributon_to_the_entanglement}
\end{equation}
Similarly we find the contribution associated with Bob's spin which is
exactly the same and the total entanglement is
\begin{equation}
E(\hat \varrho_{\sigma_A\sigma_B}) = E(\hat\varrho_{\sigma_A}) 
+ E(\hat\varrho_{\sigma_B})
= \sin^2 2\beta.
\label{eq:Entropy_rho1}
\end{equation}
It reaches its maximum value $1$ when $\beta = \pi/4$ and is zero at 
$\beta = 0, \pi/2$.

Now consider a measurement.
If $z$-component of the spin of $A$ is measured, the density matrix becomes
\begin{equation}
\hat \rho_{\sigma_{Az}\sigma_B} = 
\cos^2 \beta \ket{\uparrow\downarrow}\bra{\uparrow\downarrow} +
\sin^2 \beta \ket{\downarrow\uparrow}\bra{\downarrow\uparrow}.
\label{eq:Matrix_rho_alpha_measured}
\end{equation}
The sum of squares of the matrix elements is no longer equal to $1$,
and the entanglement is equal to
\begin{equation}
E(\hat \varrho_{\sigma_{Az}s_B}) = \frac {\sin^2 2 \beta} 2.
\label{eq:Entanglement_Simplest_sA_measured}
\end{equation}

\section{Structure of the density matrix for the boosted system}
\label{sec:BoostedMatrices}

We derive here some useful expressions for the spin density matrices $A_{ik}$, $i,k=1,2$, of Eq. (\ref{eq:FullRhoLambda}), and for the coefficients $c_i,~i=1,2,3,4$ given in Eqs. (\ref{eq:c1})-(\ref{eq:c4}). We first list some useful relations between the coefficients $c_i$:
\begin{align}
&c_1^2 + c_2^2 +c_3^2 +c_4^2 = c_1^2 +c_2^2 +2c_3^2 = 1,
\label{eq:crel1}
\\
&c_1 + c_2 = \cos \beta + \sin \beta,
\label{eq:crel2}
\\
& c_1 - c_2= \cos \delta (\cos \beta - \sin \beta),
\label{eq:crel2222}
\\
& c_1^2 + c_2 ^2 = 1- \frac 1 2 \sin^2 \delta (1-\sin 2\beta),
\label{eq:crel222}
\\
&c_1^2 + c_3^2 = \frac 1 2 (1 + \cos \delta \cos 2 \beta),
\label{eq:crel3}
\\
&c_2^2 + c_3^2 = \frac 1 2 (1 - \cos \delta \cos 2 \beta),
\label{eq:crel4}
\\
& c_1 c_2 =\frac 1 2
\left[\sin 2\beta + \frac 1 2 \sin^2 \delta (1 -\sin 2\beta)\right],
\label{eq:crel22}
\\
& c_1 c_3 = - \frac 1 4 \sin \delta \left[ \cos 2 \beta + 
\cos \delta (1-\sin 2\beta)\right],
\label{eq:crel5}
\\
& c_2 c_3 = -\frac 1 4 \sin \delta \left[ \cos 2 \beta -
\cos \delta(1-\sin 2\beta)\right].
\end{align}

Taking trace over momentum variables, we get a matrix for spins:
\begin{equation}
\hat \varrho^\Lambda_{\sigma_A\sigma_B} = 
\begin{pmatrix}
c_1^2 & c_1 c_2 & c_1 c_3 \cos 2 \alpha & c_1 c_4 \cos 2\alpha \\
c_1 c_2 & c_2^2 & c_2 c_3 \cos 2 \alpha & c_2 c_4 \cos 2 \alpha \\
c_1 c_3 \cos 2 \alpha & c_2 c_3 \cos 2 \alpha & c_3^2 & c_3 c_4 \\
c_1 c_4 \cos 2 \alpha & c_2 c_4 \cos 2 \alpha & c_3c_4 & c_4^2
\end{pmatrix}.
\label{eq:DensityMatrix_spins_boosted}
\end{equation}
Taking trace over the spin of $B$, one gets according to
Eq. (\ref{eq:Subsystem_of_Alice}):
\begin{equation}
\hat\varrho^\Lambda_{\sigma_A} =
\begin{pmatrix}
c_1^2+c_3^2 & (c_1 c_4 + c_2 c_3) \cos 2\alpha \\
(c_1 c_4 + c_2 c_3) \cos 2 \alpha & c_2^2 + c_4^2
\end{pmatrix}
\label{eq:Subsystem_of_Alice_spin_boosted}
\end{equation}
The contribution of momentum degrees of freedom to the entropy can be
calculated taking trace over spin degrees of freedom in
Eq. (\ref{eq:FullRhoLambda}) and taking into account
Eq. (\ref{eq:crel1}).
The density matrix for momenta has the form
\begin{equation}
\hat \varrho^\Lambda_{p_Ap_B} =
\begin{pmatrix}
c_\alpha^2 &  c_\alpha s_\alpha T& 0 & 0 \\
c_\alpha s_\alpha T & s_\alpha^2 & 0 & 0 \\
0 & 0 & 0 & 0 \\
0 & 0 & 0 & 0
\end{pmatrix},
\label{eq:DensityMatrix_momenta_boosted}
\end{equation}
where $c_\alpha=\cos \alpha$, $s_\alpha=\sin \alpha$,
$T=\mathrm{Tr}A_{12}= c_1^2 + c_2^2 -c_3^2 -c_4^2$ 
(note that $\mathrm{Tr}A_{11}=\mathrm{Tr}A_{22} = 1$ according
to Eq. (\ref{eq:crel1})).

If we take trace over the spin and momentum of a particle $B$, we 
get
\begin{equation}
\hat \varrho^\Lambda_{p_A \sigma_A} = \cos^2 \alpha A'_{11} 
\ket{\Lambda p_+}\bra{\Lambda p_+}
+ \sin^2 \alpha A'_{22} 
\ket{\Lambda p_-}\bra{\Lambda p_-},
\label{eq:Density_matrix_A_spin_momentum_Lambda}
\end{equation}
where $A'_{11}$ and $A'_{22}$ are the spin density matrices 
$A_{11}$ and $A_{22}$, correspondingly, reduced
by the spin of $B$:
\begin{align}
A'_{11} = &
\begin{pmatrix}
c_1^2 + c_3^2 & c_1c_4 + c_2c_3 \\
c_1c_4 + c_2 c_3 & c_2^2 + c_4^2
\end{pmatrix},
\label{eq:A11prime}
\\
A'_{22} = &
\begin{pmatrix}
c_1^2 +c_3^2 & -c_1c_4 -c_2c_3 \\
-c_1c_4 - c_2c_3 & c_2^2+c_4^2
\end{pmatrix}.
\label{eq:A22prime}
\end{align}
Finding the sum of squares of the element of the matrix of
Eq. (\ref{eq:Density_matrix_A_spin_momentum_Lambda}), one arrives at
\begin{equation}
E(\hat \varrho^\Lambda_{p_A \sigma_A}) = 1 - \frac 1 2 
(\sin^4 \alpha + \cos^4 \alpha) (1+\cos^2 2\beta),
\label{eq:Entanglement_A_spin_momentum_boosted_app}
\end{equation}
which coincides with 
Eq. (\ref{eq:entanglement}).

\section{Boosted density matrix after the measurement}
\label{sec:DerivationMeasurement}

If we measure a spin of particle $A$,
according to our previous considerations, after taking trace over the momentum
variable, the remaining spin density matrix is
matrix is equal to Eq. (\ref{eq:DensityMatrix_spins_boosted}), with
corresponding elements put identically to zero as in
Eq. (\ref{eq:DensityMatrix_AfterMeasurement_4x4}):
\begin{equation}
\hat \varrho^\Lambda_{\sigma_{Az}\sigma_B} =
\begin{pmatrix}
c_1^2 & 0 & c_1 c_3 \cos 2\alpha & 0 \\
0 & c_2^2 & 0 & c_2 c_4 \cos 2 \alpha \\
c_1 c_3 \cos 2\alpha & 0 & c_3^2 & 0 \\
0 & c_2 c_4 \cos 2\alpha & 0 & c_4^2
\end{pmatrix},
\label{eq:DensityMatrix_sAsB_spins_boosted}
\end{equation}
which after simplification by the spin of $B$ becomes
\begin{equation}
\hat \rho^\Lambda_{\sigma_{Az}} = 
\begin{pmatrix}
c_1^2 + c_3^2 & 0 \\
0 & c_2^2 +c_4^2
\end{pmatrix},
\label{eq:Subsystem_of_Alice_measured}
\end{equation}
and after taking trace by the spin of $A$,
\begin{equation}
\hat \rho^\Lambda_{\sigma_B} =
\begin{pmatrix}
c_2^2 + c_3^2 && (c_1c_3 + c_2c_4)\cos 2\alpha \\
(c_1c_3 + c_2 c_4)\cos 2 \alpha && c_1^2 +c_4^2
\end{pmatrix}.
\end{equation}

\bibliography{Quark}
\end{document}